\documentclass[12pt]{article}
\usepackage{amssymb}
\usepackage{amsmath}
\usepackage{bm}
\usepackage[dvips]{epsfig}
\usepackage{graphicx}

\begin{document}

\title{On the canonical forms of the multi-dimensional averaged
Poisson brackets.}

\author{A.Ya. Maltsev.}

\date{
\centerline{\it L.D. Landau Institute for Theoretical Physics}
\centerline{\it 142432 Chernogolovka, pr. Ak. Semenova 1A,
maltsev@itp.ac.ru}}

\maketitle

\begin{abstract}
 We consider here special Poisson brackets given by the ``averaging''
of local multi-dimensional Poisson brackets in the Whitham method.
For the brackets of this kind it is natural to ask about their
canonical forms, which can be obtained after transformations preserving
the ``physical meaning'' of the field variables. We show here that the
averaged bracket can always be written in the canonical form after
a transformation of ``Hydrodynamic Type'' in the case of absence of
annihilators of initial bracket. However, in general case the situation
is more complicated. As we show here, in more general case the averaged
bracket can be transformed to a ``pseudo-canonical'' form under some 
special (``physical'') requirements on the initial bracket.
\end{abstract}

\section{Introduction.}

 We will consider here the Poisson brackets obtained by the
``averaging'' of local multi-dimensional Poisson brackets
$$\{ \varphi^{i} ({\bf x}) \, , \, \varphi^{j} ({\bf y}) \} 
\,\,\, = \,\,\, \sum_{l_{1},\dots,l_{d}} B^{ij}_{(l_{1},\dots,l_{d})}
(\bm{\varphi}, \bm{\varphi}_{\bf x}, \dots ) \,\,\,
\delta^{(l_{1})} (x^{1} - y^{1}) \, \dots \,
\delta^{(l_{d})} (x^{d} - y^{d}) \,\,\, \equiv $$
\begin{equation}
\label{MultDimPBr}
\equiv \,\,\, \sum_{l_{1},\dots,l_{d}} B^{ij}_{(l_{1},\dots,l_{d})}
(\bm{\varphi}, \bm{\varphi}_{\bf x}, \dots ) \,\,\,
\delta_{l_{1} x^{1} \dots \, l_{d} x^{d}} \,
({\bf x} - {\bf y}) \,\,\,\,\, , \,\,\,\,\,\,\,
(l_{1}, \dots, l_{d} \geq 0)
\end{equation}
on the families of $m$-phase quasiperiodic solutions of local 
Hamiltonian systems
\begin{equation}
\label{EvInSyst}
\varphi^{i}_{t} \,\,\, = \,\,\, 
F^{i} (\bm{\varphi}, \bm{\varphi}_{\bf x},
\bm{\varphi}_{\bf xx}, \dots ) \,\,\, \equiv \,\,\,
F^{i}(\bm{\varphi}, \bm{\varphi}_{x^{1}},
\dots, \bm{\varphi}_{x^{d}}, \dots ) \,\,\,\,\, , \,\,\,\,\,\,\,\,
i = 1, \dots, n  \,\,\, ,
\end{equation}
which are represented in the following general form
\begin{equation}
\label{mphasesol}
\varphi^{i} ({\bf x}, \, t) \,\,\,\,\, = \,\,\,\,\, 
\varphi^{i}_{[{\bf a}, \bm{\theta}_{0}]} ({\bf x}, \, t) 
\,\,\,\,\, = \,\,\,\,\, \Phi^{i} \left( \, 
{\bf k}_{1}({\bf a})\, x^{1} \, + \, \dots \, + \,
{\bf k}_{d}({\bf a})\, x^{d}
\, + \, \bm{\omega}({\bf a})\, t \, + \, \bm{\theta}_{0}, \,\,
{\bf a} \, \right)
\end{equation}
with some smooth $2\pi$-periodic in each $\theta^{\alpha}$ functions
$\, \Phi^{i} (\bm{\theta}, \, {\bf a})$.

 Thus, we assume that $\, {\bf x} \, = \, (x^{1}, \dots, x^{d})$,
$\,\, {\bf y} \, = \, (y^{1}, \dots, y^{d}) \, $ represent points 
of the Euclidean space $\, \mathbb{R}^{d}$ and the expression 
(\ref{MultDimPBr}) defines a skew-symmetric Hamiltonian operator
on the space of smooth functions 
$$\bm{\varphi} ({\bf x}) \,\, = \,\, \left( \varphi^{1} 
({\bf x}) , \dots, \varphi^{n} ({\bf x}) \right)  \,\,\, , $$
satisfying the Jacobi identity.

 The notations $\, \delta^{(l)} (x - y) \, $ mean here the higher
derivatives of the delta-function and we assume that the sum in 
(\ref{MultDimPBr}) contains a finite number of terms.

 We will call brackets (\ref{MultDimPBr}) general local 
field-theoretic Poisson brackets in $\, \mathbb{R}^{d}$ and assume
that system (\ref{EvInSyst}) represents a Hamiltonian system
generated by a local Hamiltonian functional
\begin{equation}
\label{MultDimHamFunc}
H \,\, = \,\, \int P_{H} \left(\bm{\varphi}, \bm{\varphi}_{\bf x},
\bm{\varphi}_{\bf xx}, \dots \right) \,\, d^{d} x
\end{equation}
according to bracket (\ref{MultDimPBr}).

 We assume that the family (\ref{mphasesol}) is defined with the aid 
of a smooth finite-parametric set $\, {\hat \Lambda} \, $ of
$2\pi$-periodic in each $\theta^{\alpha}$ functions
$$\Phi^{i} \left( \bm{\theta} + \bm{\theta}_{0}, \, {\bf a} \right)
\,\, = \,\, \Phi^{i} \left( \theta^{1} + \theta^{1}_{0}, \dots,
\theta^{m} + \theta^{m}_{0}, \,\, a^{1} , \dots , a^{N} \right) $$
with a smooth dependence of the wave numbers
$\, {\bf k}_{q} ({\bf a}) \, = \, (k^{1}_{q} ({\bf a}), \dots ,
k^{m}_{q} ({\bf a})) \, $ and frequencies    \linebreak
$\, \bm{\omega} ({\bf a}) \, = \, (\omega^{1} ({\bf a}), \dots ,
\omega^{m} ({\bf a})) \, $ on the parameters
$\, {\bf a} \, = \, (a^{1} , \dots , a^{N})$. All the functions
$\, \Phi^{i} (\bm{\theta}, \, {\bf a} ) \, $ should satisfy the
system
\begin{equation}
\label{EvPhaseSyst}
\omega^{\alpha} \, \Phi^{i}_{\theta^{\alpha}} \,\, - \,\,
F^{i} \left( \bm{\Phi}, \,
k^{\beta_{1}}_{1} \, \bm{\Phi}_{\theta^{\beta_{1}}}, \dots,
k^{\beta_{d}}_{d} \, \bm{\Phi}_{\theta^{\beta_{d}}},
\dots \right) \,\,\, = \,\,\, 0
\end{equation}

 The parameters $\theta_{0}^{\alpha}$ represent the initial phase 
shifts of solutions (\ref{mphasesol}) and take by definition all
possible real values on the family $\, {\hat \Lambda}$. We assume
also that the values of the parameters $\, {\bf a}$ do not change
under the initial phase shifts. Let us denote by $\, \Lambda \, $
the family (\ref{mphasesol}) of the functions
$\, \varphi^{i} ({\bf x}, \, t) \, $ corresponding to the family
$\, {\hat \Lambda}$.

 The procedure of averaging of a Poisson bracket is closely connected
with the Whitham averaging method (\cite{whith1, whith2, whith3}).
For this reason we will put here additional requirements of regularity
and completeness on the family $\, \Lambda \, $ which we formulate 
below.

 Let us say first that we will everywhere consider here the generic 
situation where the values 
$\, ({\bf k}_{1}, \dots, {\bf k}_{d}, \, \bm{\omega} ) \, $
represent independent parameters on the full family of $m$-phase
solutions of system (\ref{EvInSyst}). Thus, we assume that the number
of real parameters $\, (a^{1} , \dots , a^{N}) \, $ 
is equal to   \linebreak
$\, m d \, + \, m \, + \, s \, $, $\,\, s \, \geq \, 0$.
In particular, the parameters $\, (a^{1} , \dots , a^{N}) \, $ 
can be locally chosen in the form
$\, {\bf a} \, = \, ({\bf k}_{1}, \dots, {\bf k}_{d}, \, \bm{\omega},
\, {\bf n}) \, $ where 
$\, ({\bf k}_{1}, \dots, {\bf k}_{d}, \, \bm{\omega} ) \, $
represent the wave numbers and the frequencies of the $m$-phase
solutions and $\, {\bf n} \, = \, (n^{1} , \dots , n^{s}) \, $ 
are some additional parameters (if any).

 Let us consider now linear operators
$\, {\hat L}^{i}_{j [{\bf a}, \bm{\theta}_{0}]} \, = \,
{\hat L}^{i}_{j [{\bf k}_{1}, \dots, {\bf k}_{d}, \bm{\omega},
{\bf n}, \bm{\theta}_{0}]} \, $ given by the linearization of system
(\ref{EvPhaseSyst}) on the corresponding solutions
$\bm{\Phi} ( \bm{\theta} + \bm{\theta}_{0}, \, 
{\bf k}_{1}, \dots, {\bf k}_{d}, \, \bm{\omega},\, {\bf n}) \, $.
It's not difficult to see that the functions
$\bm{\Phi}_{\theta^{\alpha}} ( \bm{\theta} + \bm{\theta}_{0}, \,
{\bf k}_{1}, \dots, {\bf k}_{d}, \, \bm{\omega},\, {\bf n})$,
$\,\, \alpha \, = \, 1, \dots, m$, $\,\, $ and
$\,\, \bm{\Phi}_{n^{l}} ( \bm{\theta} + \bm{\theta}_{0}, \,
{\bf k}_{1}, \dots, {\bf k}_{d}, \, \bm{\omega},\, {\bf n})$,   
$\,\, l \, = \, 1, \dots, s$, $\,\, $ represent kernel vectors of 
the operators 
$\, {\hat L}^{i}_{j [{\bf k}_{1}, \dots, {\bf k}_{d}, \bm{\omega},
{\bf n}, \bm{\theta}_{0}]} \, $ on the space of $2\pi$-periodic
in each $\theta^{\alpha}$ functions which depend smoothly on all
the parameters 
$\, ({\bf k}_{1}, \dots, {\bf k}_{d}, \, \bm{\omega},\, {\bf n}, 
\, \bm{\theta}_{0}) $. Let us put now the following requirements
on the operators 
$\, {\hat L}^{i}_{j [{\bf k}_{1}, \dots, {\bf k}_{d}, \bm{\omega},
{\bf n}, \bm{\theta}_{0}]} \, $ on the family $\, {\hat \Lambda} \, $:

\vspace{0.2cm}

1) We require that the vectors
$\, \bm{\Phi}_{\theta^{\alpha}} (\bm{\theta} + \bm{\theta}_{0}, \,
{\bf k}_{1}, \dots, {\bf k}_{d}, \bm{\omega}, {\bf n})$,
$\, \bm{\Phi}_{n^{l}} (\bm{\theta} + \bm{\theta}_{0}, \,
{\bf k}_{1}, \dots, {\bf k}_{d}, \bm{\omega}, {\bf n}) \, $   
are linearly independent and represent the maximal linearly 
independent set among the kernel vectors of the operator 
$\, {\hat L}^{i}_{j[{\bf k}_{1}, \dots, {\bf k}_{d},
\bm{\omega},{\bf n},\bm{\theta}_{0}]}$ 
on the space of $2\pi$-periodic in each $\theta^{\alpha}$ functions
smoothly depending on the parameters
$({\bf k}_{1}, \dots, {\bf k}_{d}, \, \bm{\omega}, \, {\bf n})$.

\vspace{0.2cm}

2) The operators
$\, {\hat L}^{i}_{j[{\bf k}_{1}, \dots, {\bf k}_{d},
\bm{\omega},{\bf n},\bm{\theta}_{0}]} \, $ have exactly 
$\, m \, + \, s \, $ linearly independent regular left eigen-vectors
$\bm{\kappa}^{(q)}_{[{\bf k}_{1}, \dots, {\bf k}_{d},
\bm{\omega}, {\bf n}]} (\bm{\theta} + \bm{\theta}_{0})$, 
$\,\, q = 1, \dots , m + s $, on the space of $2\pi$-periodic in each 
$\theta^{\alpha}$ functions, corresponding to the zero eigenvalue and
depending smoothly on the parameters
$({\bf k}_{1}, \dots, {\bf k}_{d}, \, \bm{\omega}, \, {\bf n})$.
 
\vspace{0.3cm}

\noindent
{\bf Definition 1.1.}

{\it Under all the requirements formulated above we will call the 
corresponding family $\, \Lambda \, $ a {\bf complete regular family}
of $m$-phase solutions of system (\ref{EvInSyst}).
}

\vspace{0.3cm}

 It is well known that the Whitham approach gives a description of the
slowly modulated $m$-phase solutions of nonlinear PDE's. The Whitham
solutions represent asymptotic solutions of nonlinear systems with the
main part having the form
\begin{equation}
\label{MainApprWhith}
\bm{\varphi}_{(0)} \, ({\bf x}, \, t, \, \bm{\theta}) \,\,\, = \,\,\,
\bm{\Phi} \left( {{\bf S} ({\bf X}, T) \over \epsilon} \, + \,
\bm{\theta}_{(0)} ({\bf X}, T) \, + \, \bm{\theta}, \,\,
{\bf S}_{X^{1}}, \dots, {\bf S}_{X^{d}}, \, {\bf S}_{T}, \,
{\bf n} ({\bf X}, T) \right)
\end{equation}
where $\, {\bf X} \, = \, \epsilon \, {\bf x}$,
$\, T \, = \, \epsilon \, t$, $\, \epsilon \rightarrow 0$, $\,$
are the slow spatial and time variables and the function
$${\bf S} ({\bf X}, T) \,\, = \,\, \left(
S^{1} ({\bf X}, T), \dots, S^{m} ({\bf X}, T) \right) $$
represents the ``modulated phase'' of the solution. Thus, the main part
of the Whitham solution represents an $m$-phase solution of the
nonlinear system with the slow modulated parameters 
$\, {\bf a} ({\bf X}, T) \, $ and a rapidly changing phase. 
We have also the natural connection
\begin{equation}
\label{SOmKRel}
S^{\alpha}_{T} \,\, = \,\, \omega^{\alpha} ({\bf X}, T) \,\,\, ,
\,\,\,\,\,\,\,\, 
S^{\alpha}_{X^{q}} \,\, = \,\, k^{\alpha}_{q} ({\bf X}, T)
\end{equation}
between the derivatives of the modulated phase and the parameters
$\, \bm{\omega} ({\bf X}, T) \, $ and $\, {\bf k}_{q} ({\bf X}, T)$.

 Relations (\ref{SOmKRel}) give the natural constraints
$$k^{\alpha}_{q T} \,\, = \,\, \omega^{\alpha}_{X^{q}} 
\,\,\, , \,\,\,\,\,\,\,\,
k^{\alpha}_{q X^{p}} \,\, = \,\, k^{\alpha}_{p X^{q}} $$
on the functions $\, \bm{\omega} ({\bf X}, T) \, $ and 
$\, {\bf k}_{q} ({\bf X}, T) $, which can be considered as the first
part of the Whitham system on the parameters $\, {\bf a} ({\bf X}, T)$.

 The second part of the Whitham system is defined usually by the
requirement of existence of a bounded next correction to the initial
approximation (\ref{MainApprWhith}) and can be defined in different
ways which are usually equivalent to each other (see e.g.
\cite{whith1, whith2, whith3, luke, ffm, dm, Haberman1, Haberman2,
dn2, dn3, krichev1}).

 In our scheme we will define the second part of the Whitham system
for a complete regular family $\, \Lambda \, $ of $m$-phase solutions
of (\ref{EvInSyst}) as the orthogonality at every ${\bf X}$ and $T$
of all the regular left eigen-vectors
$$\bm{\kappa}^{(q)}_{[{\bf S}_{X^{1}}, \dots, {\bf S}_{X^{d}}, 
{\bf S}_{T}, {\bf n}({\bf X}, T)]} \left( 
{{\bf S} ({\bf X}, T) \over \epsilon} \, + \,
\bm{\theta}_{(0)} ({\bf X}, T) \, + \, \bm{\theta} \right) 
\,\,\,\,\, , \,\,\,\,\,\,\,\, q = 1, \dots, m + s $$
to the first $\epsilon$-discrepancy 
$\, {\bf f}_{1} (\bm{\theta}, {\bf X}, T)$, obtained after the
substitution of the main approximation (\ref{MainApprWhith}) into
the system
$$\epsilon \, \varphi^{i}_{T} \,\,\, = \,\,\,
F^{i} \left( \bm{\varphi}, \, \epsilon \, \bm{\varphi}_{\bf X},
\, \epsilon^{2} \, \bm{\varphi}_{\bf XX}, \, \dots \right) $$

 It is well known that the full Whitham system, defined in one of the
standard ways, does not put any restrictions on the variables
$\, \bm{\theta}_{0} ({\bf X}, T) \, $ and represents a system of
PDE's just on the parameters $\, {\bf a} ({\bf X}, T) \, $
(see e.g. \cite{whith1, whith2, whith3, luke}). In particular, it is
also not difficult to show that the orthogonality conditions
\begin{equation}
\label{KappafOrtRel}
\int_{0}^{2\pi}\!\!\!\!\!\dots\int_{0}^{2\pi} \! 
\kappa^{(q)}_{[{\bf S}_{X^{1}}, \dots, {\bf S}_{X^{d}},
{\bf S}_{T}, {\bf n}({\bf X}, T)] \, i} \left(
{{\bf S} ({\bf X}, T) \over \epsilon} \, + \,
\bm{\theta}_{(0)} ({\bf X}, T) \, + \, \bm{\theta} \right) \,
f^{i}_{1} (\bm{\theta}, {\bf X}, T) \,\,
{d^{m} \theta \over (2\pi)^{m}} \,\,\, = \,\,\, 0
\end{equation}
defined for any complete regular family $\Lambda$, possesses
the same property (see e.g. \cite{Sigma, JMP2}
\footnote{The definition of ${\bf f}_{1}$ in \cite{Sigma, JMP2}
differs by a phase shift from that used here which is included there 
also in the corresponding orthogonality conditions.}) In general,
relations (\ref{KappafOrtRel}) can be written as a system of
$\, m + s \, $ quasilinear equations
$$P^{(q)}_{\alpha} \left( {\bf S}_{\bf X}, {\bf S}_{T}, {\bf n} \right)
\, S^{\alpha}_{TT} \,\,\, + \,\,\, 
Q^{(q)p}_{\alpha} \left( {\bf S}_{\bf X}, {\bf S}_{T}, {\bf n} \right)
\, S^{\alpha}_{X^{p} T} \,\,\, + \,\,\,
R^{(q)pk}_{\alpha} \left( {\bf S}_{\bf X}, {\bf S}_{T}, {\bf n} \right)
\, S^{\alpha}_{X^{p} X^{k}} \,\,\, +  $$
$$ + \,\,\, V^{(q)}_{l} 
\left( {\bf S}_{\bf X}, {\bf S}_{T}, {\bf n} \right)
\, n^{l}_{T} \,\,\, + \,\,\, 
W^{(q)p}_{l} \left( {\bf S}_{\bf X}, {\bf S}_{T}, {\bf n} \right)
\, n^{l}_{X^{P}} \,\,\,\, = \,\,\, 0 \,\,\,\,\, , \,\,\,\,\,\,\,\,\,\, 
q \, = \, 1, \dots, m + s $$
with some smooth functions
$\, P^{(q)}_{\alpha}$, $\, Q^{(q)p}_{\alpha}$, $\, R^{(q)pk}_{\alpha}$,
$\, V^{(q)}_{l}$, $\, W^{(q)p}_{l}$.

 Let us say here that for the single-phase case ($m = 1$) the set of 
the ``regular'' left eigen-vectors 
$\, \bm{\kappa}^{(q)}_{[k_{1}, \dots, k_{d},
\omega, {\bf n}]} (\theta + \theta_{0})$,
$\,\, q = 1, \dots, s + 1$, $\, $ represents usually the full set of 
linearly independent left eigen-vectors of the operators
$\, {\hat L}^{i}_{j [k_{1}, \dots, k_{d}, \omega, {\bf n}, \theta_{0}]}$,
corresponding to the zero eigen-value, for all the values of
$\, (k_{1}, \dots, k_{d}, \omega, {\bf n}, \theta_{0})$ on a complete
regular family $\, \Lambda$. However, for the multi-phase 
case    \linebreak
($m > 1$) the situation is usually more complicated 
and ``irregular'' left eigen-vectors of 
$\, {\hat L}^{i}_{j[{\bf k}_{1}, \dots, {\bf k}_{d},
\bm{\omega},{\bf n},\bm{\theta}_{0}]}$, 
corresponding to the zero eigen-value, also arise for
special values of parameters
$\, ({\bf k}_{1}, \dots, {\bf k}_{d}, \, \bm{\omega}, \, {\bf n})$.
As a result, the corrections to the main approximation
(\ref{MainApprWhith}) of the Whitham solution for the multi-phase
case have usually rather different form in comparison with the case
$m = 1$ (see e.g. \cite{dobr1, dobr2, DobrKrichever}).

 Let us say, however, that the regular Whitham system still plays 
the central role in the description of the slow-modulated $m$-phase
solutions both in the cases $\, m = 1 \, $ and $\, m > 1 $. Let us
give here also just some incomplete list of classical papers devoted
to different questions connected with the Whitham approach:
\cite{AblBenny, dm, DobrMaslMFA, dobr1, dobr2, DobrKrichever,
dn1, dn2, dn3, DubrovinCMP, ffm, Hayes, krichev1, KricheverCPAM,
luke, Newell, theorsol, Nov, whith1, whith2, whith3}.

 One of the most elegant ways of constructing the Whitham system
was suggested by Whitham and is connected with the averaging of the
Lagrangian function of the initial system. This method is applicable
to any system having a local Lagrangian structure and gives a local 
Lagrangian structure for the corresponding Whitham system (see e.g.
\cite{whith3}). Let us say, that the Lagrangian approach gives usually
essential advantages both in constructing and investigation of the
Whitham equations.

 The class of local Lagrangian systems can be significantly expanded
being included into a larger class of systems having local
field-theoretic Hamiltonian structure. In general, the systems of this 
kind can be considered as the evolution systems (\ref{EvInSyst})
which can be represented in the form
$$\varphi^{i}_{t} \,\,\, = \,\,\, {\hat J}^{ij} \,\,
{\delta H \over \delta \varphi^{j} ({\bf x})} $$
where $\, {\hat J}^{ij} \, $ is the Hamiltonian operator
$${\hat J}^{ij} \,\,\, = \,\,\,
\sum_{l_{1},\dots,l_{d}} B^{ij}_{(l_{1},\dots,l_{d})}
(\bm{\varphi}, \bm{\varphi}_{\bf x}, \dots ) \,\,
\left({d \over d x^{1}} \right)^{l_{1}} \, \dots \,\,
\left({d \over d x^{d}} \right)^{l_{d}} \,\,\, , $$
defined by the Poisson bracket (\ref{MultDimPBr}), and $H$ is the 
Hamiltonian functional having the form (\ref{MultDimHamFunc}).

 The Hamiltonian theory of the Whitham equations was started by
B.A. Dubrovin and S.P. Novikov, who introduced the concept of the
Hamiltonian structure of Hydrodynamic Type. In general, the
Dubrovin-Novikov bracket in $\mathbb{R}^{d}$ can be written in the 
following local form
\begin{equation}
\label{DNbracket}
\left\{ U^{\nu} ({\bf X}) \, , \, U^{\mu} ({\bf Y}) \right\} 
\,\,\, = \,\,\, g^{\nu\mu \, l} \left( {\bf U}({\bf X}) \right) \,
\delta_{X^{l}} ( {\bf X} - {\bf Y} ) \,\, + \,\,
b^{\nu\mu \, l}_{\lambda} \left( {\bf U}({\bf X}) \right) \,
U^{\lambda}_{X^{l}} \,\, \delta ( {\bf X} - {\bf Y} )
\end{equation}
(summation over repeated indices).

 The general theory of the brackets (\ref{DNbracket}) is rather 
nontrivial. Rather deep results on the classification of
brackets (\ref{DNbracket}) were obtained in 
\cite{DubrNovDAN84,MokhovMultDimBr1,MokhovMultDimBr2} where the full
description of brackets (\ref{DNbracket}), satisfying special
non-degeneracy conditions, was presented. However, there are many
interesting examples where a nontrivial structure of a system is
defined by a non-generic bracket (\ref{DNbracket}) 
(see e.g. \cite{FerOdesSt, FerLorSav}). In general, we can say that the 
full theory of the brackets (\ref{DNbracket}) represents an important 
branch of the theory of the Poisson brackets and is still waiting 
for its final completion. 

 A special class of the Dubrovin-Novikov brackets (\ref{DNbracket})
is given by the one-dimensional brackets of Hydrodynamic Type.
The brackets (\ref{DNbracket}) have in this case the following
general form
\begin{equation}
\label{OneDimDNBracket}
\{U^{\nu}(X) \, , \, U^{\mu}(Y)\} \,\,\, = \,\,\, 
g^{\nu\mu}({\bf U}(X)) \,\, \delta^{\prime}(X-Y) \,\,\, + \,\,\,
b^{\nu\mu}_{\lambda}({\bf U}(X) ) \,\, U^{\lambda}_{X} \,\,\, 
\delta (X-Y)  \quad  ,  \quad \,\,\,  \nu , \mu = 1, \dots , N
\end{equation}
and are closely connected with Differential Geometry. Thus, it can
be proved (\cite{dn1,DubrNovDAN84,dn2,dn3}) that the expression
(\ref{OneDimDNBracket}) with non-degenerate
tensor $\, g^{\nu\mu}({\bf U})$ defines a Poisson bracket on the 
space of fields $\, {\bf U}(X)$ if and only if the tensor 
$\, g^{\nu\mu}({\bf U})$ defines a flat pseudo-Riemannian metric
with upper indices on the space of $\, {\bf U}$ while the values
$\, \Gamma^{\nu}_{\mu\gamma} \, = \, - \, g_{\mu\lambda} \,
b^{\lambda\nu}_{\gamma} \, $ represent the corresponding Christoffel 
symbols ($g_{\nu\tau}({\bf U}) \, g^{\tau\mu}({\bf U}) 
\, = \, \delta^{\mu}_{\nu}$).

 As a consequence, we can claim in fact that every Poisson bracket
(\ref{OneDimDNBracket}) with non-degenerate tensor 
$\, g^{\nu\mu}({\bf U}) \, $ can be locally written in the constant
form 
\begin{equation}
\label{DNBrConstForm}
\{n^{\nu}(X) \, , \, n^{\mu}(Y)\} \,\,\, = \,\,\, \epsilon^{\nu}
\, \delta^{\nu\mu} \,\, \delta^{\prime}(X-Y) \,\,\, , \,\,\,\,\,\,\,\,
\epsilon^{\nu} = \pm 1 
\end{equation}
after the transition to the flat coordinates
$\, n^{\nu} \, = \, n^{\nu}({\bf U}) \, $ 
of the metric $\, g_{\nu\mu}({\bf U}) $.

 It's not difficult to see also, that the functionals
$$N^{\nu} \,\,\, = \,\,\, \int_{-\infty}^{+\infty} 
n^{\nu}(X) \,\, dX $$
represent then the annihilators of the bracket
(\ref{OneDimDNBracket}) while the functional
$$P \,\,\, = \,\,\, \int_{-\infty}^{+\infty} \, {1 \over 2} \,
\sum_{\nu=1}^{N} \, \epsilon^{\nu}  \, (n^{\nu})^{2} (X) \,\,\, dX$$
gives the momentum operator for the bracket (\ref{OneDimDNBracket})
$\, $ (\cite{dn1,DubrNovDAN84,dn2,dn3}).

 The statement, formulated above, plays in fact the role of an analog 
of the Darboux Theorem for the brackets (\ref{OneDimDNBracket}) with
non-degenerate tensor $\, g_{\nu\mu}({\bf U}) $. Following
B.A. Dubrovin and S.P. Novikov, we will call the form
(\ref{DNBrConstForm}) of the bracket (\ref{OneDimDNBracket}) the
Canonical Form of a non-degenerate one-dimensional Poisson bracket
of Hydrodynamic type. Let us note here also, that the theory of
the brackets (\ref{OneDimDNBracket}) with degenerate tensor 
$\, g_{\nu\mu}({\bf U}) \, $ can be also formulated
in a nice Differential Geometric form which we will not consider 
here in detail (\cite{Grinberg}).

 The theory of the Poisson brackets of Hydrodynamic Type gives the
basement for the theory of integrability of multi-component 
one-dimensional Hydrodynamic Type systems 
\begin{equation}
\label{OneDimHTSyst}
U^{\nu}_{T} \,\,\, = \,\,\, V^{\nu}_{\mu} ({\bf U}) \,\,
U^{\mu}_{X} \,\,\,\,\, , \,\,\,\,\,\,\,\,
\nu = 1, \dots, N
\end{equation}

 Thus, according to conjecture of S.P. Novikov, every diagonalizable
system (\ref{OneDimHTSyst}) which is Hamiltonian with respect to
some bracket (\ref{OneDimDNBracket}) with the Hamiltonian of
Hydrodynamic Type
$$H  \,\,\, = \,\,\,  \int_{-\infty}^{+\infty} \,
h({\bf U}) \,\, dX $$
can be integrated.

 The conjecture of S.P. Novikov was proved by S.P. Tsarev
(\cite{tsarev, tsarev2}) who suggested a method for solving of
diagonal Hamiltonian systems
\begin{equation}
\label{DiagHTSyst}
U^{\nu}_{T} \,\,\, = \,\,\, V^{\nu} ({\bf U}) \,\,
U^{\nu}_{X} \,\,\,\,\, , \,\,\,\,\,\,\,\,
\nu = 1, \dots, N
\end{equation}

 The method of Tsarev can be applied in fact to a wider class of
systems (\ref{DiagHTSyst}) which were called by S.P. Tsarev
semi-Hamiltonian. In particular, the class of the semi-Hamiltonian
systems contains the diagonal systems, Hamiltonian with respect to 
the weakly nonlocal Poisson brackets of Hydrodynamic Type - the 
Mokhov-Ferapontov bracket (\cite{mohfer1}) and more general
Ferapontov brackets (\cite{fer1, fer2}), which appeared as 
generalizations of the brackets of B.A. Dubrovin and S.P. Novikov.
The diagonal semi-Hamiltonian systems represent one of the widest 
classes of integrable one-dimensional systems of Hydrodynamic Type.

 B.A. Dubrovin and S.P. Novikov suggested also a method of
averaging of local field-theoretic Hamiltonian structures for the
case of one spatial dimension.

 The Dubrovin-Novikov procedure is based on the existence of $N$
local integrals of system (\ref{EvInSyst}) 
$$I^{\nu} \,\,\, = \,\,\, \int
P^{\nu}(\bm{\varphi}, \bm{\varphi}_{x},\dots) \, dx $$
which commute with the Hamiltonian $H$ and with each other
\begin{equation}
\label{CommInt}
\{I^{\nu} \, , \, H\} \, = \, 0 \,\,\,\,\, , \,\,\,\,\,\,\,\,  
\{I^{\nu} \, , \, I^{\mu}\} \, = \, 0 
\end{equation}
according to the bracket (\ref{MultDimPBr}) $\, $ ($d = 1$). It is
supposed also that the set of parameters $\, {\bf a} \, $ on the
family $\, \Lambda \, $ can be chosen in the form
$\, (a^{1}, \dots, a^{N}) \, = \, (U^{1}, \dots, U^{N}) \, $,
where
$$U^{\nu} \,\,\, = \,\,\, \langle P^{\nu} \rangle 
\,\,\, \equiv \,\,\, \int_{0}^{2\pi}\!\!\!\dots\int_{0}^{2\pi} \,
P^{\nu} \left( \bm{\Phi}, \, k^{\alpha} \bm{\Phi}_{\theta^{\alpha}},
\dots \right) \,\, {d^{m} \theta \over (2\pi)^{m}} $$
represent the values of the densities
$\, P^{\nu}(\bm{\varphi}, \bm{\varphi}_{x},\dots) \, $ on
$\Lambda$, averaged over the angle (phase) variables.

 We can write for the time evolution of the densities
$\, P^{\nu}(\bm{\varphi}, \bm{\varphi}_{x},\dots) \, $
according to system (\ref{EvInSyst}):
$$P^{\nu}_{t} (\bm{\varphi}, \bm{\varphi}_{x},\dots)
\,\,\, \equiv \,\,\,
Q^{\nu}_{x} (\bm{\varphi}, \bm{\varphi}_{x},\dots) \,\,\, , $$
where $\, Q^{\nu} (\bm{\varphi}, \bm{\varphi}_{x}, \dots) \, $
are some smooth functions of $\, \bm{\varphi} \, $ and its 
spatial derivatives. It is convenient to write also the Whitham
system as a system of conservation laws
\begin{equation}
\label{OneDimConsWhithSyst}
\langle P^{\nu} \rangle_{T} \,\,\, = \,\,\,
\langle Q^{\nu} \rangle_{X} \,\,\,\,\, , \,\,\,\,\,\,\,\,
\nu = 1, \dots, N \,\,\, ,
\end{equation}
using the functions
$\, P^{\nu}(\bm{\varphi}, \bm{\varphi}_{x},\dots) \, $ and 
$\, Q^{\nu} (\bm{\varphi}, \bm{\varphi}_{x}, \dots)$.

 The procedure of construction of the Dubrovin-Novikov bracket
for system (\ref{OneDimConsWhithSyst}) can be described in the
following way:

 Let us calculate the pairwise Poisson brackets of the
densities $\, P^{\nu} (x)$, $\, P^{\mu} (y)$, which can be
represented in the form:
$$\{P^{\nu}(x) \, , \, P^{\mu}(y)\}
\,\,\, = \,\,\, \sum_{k\geq 0} \,
A^{\nu\mu}_{k}(\bm{\varphi}, \bm{\varphi}_{x}, \dots) \,\,
\delta^{(k)}(x-y) $$
which some smooth functions 
$\, A^{\nu\mu}_{k}(\bm{\varphi}, \bm{\varphi}_{x}, \dots) $.

 According to conditions (\ref{CommInt}) we can write the relations
$$A^{\nu\mu}_{0}(\bm{\varphi}, \bm{\varphi}_{x}, \dots)
\,\,\, \equiv \,\,\,
\partial_{x} Q^{\nu\mu}(\bm{\varphi}, \bm{\varphi}_{x},
\dots) $$
for some functions
$\, Q^{\nu\mu} (\bm{\varphi}, \bm{\varphi}_{x},\dots)$.

 Let us put now $\, U^{\nu} \, = \, \langle P^{\nu} \rangle \, $
and define the Poisson bracket
\begin{equation}
\label{OneDimAvBr}
\left\{U^{\nu}(X) \, , \, U^{\mu}(Y) \right\} \,\,\,\,\, = \,\,\,\,\,   
\langle A^{\nu\mu}_{1}\rangle ({\bf U}) \,\,\, 
\delta^{\prime}(X-Y) \,\,\, + \,\,\,
{\partial \langle Q^{\nu\mu} \rangle \over
\partial U^{\gamma}} \,\, U^{\gamma}_{X} \,\,\, \delta (X-Y) 
\end{equation}
on the space of functions $\, {\bf U} ({\bf X})$.

 System (\ref{OneDimConsWhithSyst}) can be defined now as a 
Hamiltonian system with respect to the bracket (\ref{OneDimAvBr})
with the Hamiltonian functional
$$H_{av} \,\,\, = \,\,\, \int_{-\infty}^{+\infty} \, 
\langle P_{H} \rangle \left( {\bf U} (X) \right) \,\, d X $$

 Let us say that the complete justification of the Dubrovin-Novikov
procedure represents in fact a nontrivial question. Let us give
here the reference on paper \cite{Sigma} where some review of this
question and the most detailed consideration of the justification
problem were presented. In particular, we can state that the 
Dubrovin-Novikov procedure is well justified for a complete regular
family $\, \Lambda \, $ having certain regular Hamiltonian
properties (\cite{Sigma}).

 In the case of several spatial dimensions ($d \, > \, 1$) the 
procedure of bracket averaging should be actually modified, which is
connected mostly with a special role of the variables 
$\, {\bf S} ({\bf X}) \, $ revealed in this situation. Let us formulate
here the corresponding procedure and the conditions of its applicability
according to the scheme proposed in \cite{JMP2, MinimalSet}.

 Let us consider a complete regular family $\, \Lambda \, $ of
$m$-phase solutions of system (\ref{EvInSyst}) parametrized by the
$\, m (d + 1) + s \, $ parameters 
$\, ({\bf k}_{1}, \dots, {\bf k}_{d}, \, \bm{\omega}, \, {\bf n}) \, $
and $m$ initial phase shifts $\bm{\theta}_{0}$.

\vspace{0.3cm}

\noindent
{\bf Definition 1.2.}

{\it We will call a complete regular family $\, \Lambda \, $ a 
{\bf complete Hamiltonian family} of $m$-phase solutions of 
(\ref{EvInSyst}) if it satisfies the following requirements:

\vspace{0.2cm}

1) The bracket (\ref{MultDimPBr}) has at every point
$\, ({\bf k}_{1}, \dots, {\bf k}_{d}, \, \bm{\omega}, \, {\bf n},
\, \bm{\theta}_{0} ) \, $ of $\, \Lambda \, $ the same number
$\, s^{\prime} \, $ of ``annihilators'' defined by linearly independent
solutions $\, {\bf v}^{(k)}_{[{\bf a}, \bm{\theta}_{0}]} ({\bf x}) \, $
of the equation
\begin{equation}
\label{AnnSolCoordRepr}
\sum_{l_{1},\dots,l_{d}} \,
B^{ij}_{(l_{1},\dots,l_{d})} (\bm{\varphi}_{[{\bf a}, \bm{\theta}_{0}]}, 
\, \bm{\varphi}_{[{\bf a}, \bm{\theta}_{0}] \, \bf x}, \dots) \,\,\,
v^{(k)}_{ [{\bf a}, \bm{\theta}_{0}] \, j \, , 
\, l_{1} x^{1} \dots \, l_{d} x^{d}}
({\bf x}) \,\,\,\,\, = \,\,\,\,\, 0 \,\,\, , 
\end{equation}
such that all the functions 
$\, v^{(k)}_{[{\bf a}, \bm{\theta}_{0}]\, i} \, ({\bf x}) \, $ can be
represented in the form
\begin{equation}
\label{AnnVarDerForm}
v^{(k)}_{[{\bf a}, \bm{\theta}_{0}]\, i} \, ({\bf x}) \,\,\, = \,\,\,
{\hat v}^{(k)}_{[{\bf a}, \bm{\theta}_{0}]\, i} \, \left(
{\bf k}_{1} x^{1} \, + \, \dots \, + \, {\bf k}_{d} x^{d} \right) 
\end{equation}
for some smooth $2\pi$-periodic in each $\theta^{\alpha}$ functions
$\, {\hat v}^{(k)}_{[{\bf a}, \bm{\theta}_{0}]\, i} \, (\bm{\theta})$.

\vspace{0.2cm}

2) For the derivatives $\, \bm{\varphi}_{\omega^{\alpha}}$,
$\, \bm{\varphi}_{n^{l}} \, $ of the functions
$\, \bm{\varphi}_{[{\bf a}, \bm{\theta}_{0}]} \, ({\bf x}) \, = \,
\bm{\varphi}_{[{\bf k}_{1}, \dots, {\bf k}_{d}, \, \bm{\omega}, 
\, {\bf n}, \, \bm{\theta}_{0}]} \, ({\bf x}) \, $ we have the
relations
$${\rm rank} \,
\begin{Vmatrix}
(\bm{\varphi}_{\omega^{\alpha}} \, \cdot \,
{\bf v}^{(k)} )  \\
(\bm{\varphi}_{n^{l}} \, \cdot \,
{\bf v}^{(k)} )
\end{Vmatrix} \,\,\, = \,\,\, s^{\prime} $$
($\alpha = 1, \dots, m$, $\, l = 1, \dots, s$,
$\, k = 1, \dots, s^{\prime}$), $ \, $ where the expressions
$$\left( \bm{\varphi}_{\omega^{\alpha}} \, \cdot \,
{\bf v}^{(k)} \right) \,\,\,\,\, \equiv \,\,\,\,\, 
\lim_{K\rightarrow\infty} \,\,\,
{1 \over (2K)^{d}} \,\, \int_{-K}^{K} \! \dots \int_{-K}^{K} \,\,
\varphi^{i}_{\omega^{\alpha}} ({\bf x}) \,\, 
v^{(k)}_{i} ({\bf x}) \,\, d^{d} x $$
$$\left( \bm{\varphi}_{n^{l}} \, \cdot \,
{\bf v}^{(k)} \right) \,\,\,\,\, \equiv \,\,\,\,\, 
\lim_{K\rightarrow\infty} \,\,\,
{1 \over (2K)^{d}} \,\, \int_{-K}^{K} \! \dots \int_{-K}^{K} \,\,
\varphi^{i}_{n^{l}} ({\bf x}) \,\,
v^{(k)}_{i} ({\bf x}) \,\, d^{d} x $$
represent the convolutions of the variation derivatives of
annihilators with the tangent vectors
$\bm{\varphi}_{\omega^{\alpha}}$,
$\, \bm{\varphi}_{n^{l}}$.
}

\vspace{0.3cm}

 It is convenient to introduce here also the families
$\, \Lambda_{{\bf k}_{1}, \dots, {\bf k}_{d}} \, $
representing the functions   \linebreak
$\, \bm{\varphi}_{[{\bf k}_{1}, \dots, {\bf k}_{d}, \, \bm{\omega},
\, {\bf n}, \, \bm{\theta}_{0}]} \, $ with the fixed parameters
$\, ({\bf k}_{1}, \dots, {\bf k}_{d})$. Following \cite{MinimalSet},
we will give here the following definition:

\vspace{0.3cm}

\noindent
{\bf Definition 1.3.}

{\it We say that a complete Hamiltonian family
$\, \Lambda \, $ is {\bf equipped with a minimal
set of commuting integrals} if there exist $m + s$ functionals
$\, I^{\gamma}$, $\,\, \gamma = 1, \dots, m + s$, $ \, $ having 
the form
\begin{equation}
\label{IgammaForm}
I^{\gamma} \,\,\, = \,\,\, \int
P^{\gamma} \left( \bm{\varphi}, \, \bm{\varphi}_{\bf x}, \,
\bm{\varphi}_{\bf xx}, \dots \right) \, d^{d} x
\end{equation}
such that:

\vspace{0.2cm}

1) The functionals $\, I^{\gamma} \, $ commute with the Hamiltonian
functional (\ref{MultDimHamFunc}) and with each other according to the
bracket (\ref{MultDimPBr}):
\begin{equation}
\label{CommutativeSet}
\left\{ I^{\gamma} \, , \, H \right\} \,\, = \,\, 0 \,\,\,\,\, ,
\,\,\,\,\,\,\,\, \left\{ I^{\gamma} \, , \, I^{\rho} \right\}
\,\, = \,\, 0 \,\,\, ,
\end{equation}

\vspace{0.2cm}

2) The values $ \, U^{\gamma}$: 
$$U^{\gamma} \,\,\,\,\, = \,\,\,\,\, \lim_{K\rightarrow\infty} \,\,\,
{1 \over (2K)^{d}} \,\, \int_{-K}^{K} \! \dots \int_{-K}^{K} \,\,
P^{\gamma} \left( \bm{\varphi}_{[{\bf a}, \bm{\theta}_{0}]},
\, \bm{\varphi}_{[{\bf a}, \bm{\theta}_{0}] \, \bf x}, \dots \right)
\,\, d^{d} x $$
of the functionals 
$ \, I^{\gamma} \, $ on $\, \Lambda \, $ represent independent 
parameters on every family
$\, \Lambda_{{\bf k}_{1}, \dots, {\bf k}_{d}}$, such that the total 
set of parameters on $\, \Lambda \, $ can be represented in the 
form $\, ({\bf k}_{1}, \dots, {\bf k}_{d}, \, U^{1}, \dots, U^{m+s}, 
\, \bm{\theta}_{0})$;

\vspace{0.2cm}

3) The Hamiltonian flows, generated by the functionals $\, I^{\gamma}$,
leave invariant the family $\, \Lambda \, $ and the values of all the 
parameters $\, ({\bf k}_{1}, \dots, {\bf k}_{d}, \, {\bf U}) \, $
of the functions $\, \bm{\varphi}_{[{\bf k}_{1}, \dots, {\bf k}_{d},
\, {\bf U}, \, \bm{\theta}_{0}]} \, ({\bf x}) \, $ 
and generate the linear time evolution of the phase shifts 
$\bm{\theta}_{0}$ with constant
frequencies
$\, \bm{\omega}^{\gamma} =
(\omega^{1\gamma}, \dots, \omega^{m\gamma})$,
such that 
\begin{equation}
\label{RankProperty}
{\rm rk} \,\, \left| \left| \, \omega^{\alpha \gamma} \, 
({\bf k}_{1}, \dots, {\bf k}_{d}, \, {\bf U}) \, \right| \right|
\,\,\, = \,\,\, m 
\end{equation}
everywhere on $\, \Lambda$;

\vspace{0.2cm}

4) At every point
$\, ({\bf k}_{1}, \dots, {\bf k}_{d}, \, {\bf U}, \, 
\bm{\theta}_{0}) \, $ of $\, \Lambda \, $
the linear space, generated by the variation derivatives
$\, \delta I^{\gamma} / \delta \varphi^{i} ({\bf x})$, contains  
the variation derivatives 
$\, {\bf v}^{(k)}_{[{\bf k}_{1}, \dots, {\bf k}_{d}, {\bf U}, 
\bm{\theta}_{0}]} ({\bf x}) \, $
of all the annihilators of bracket (\ref{MultDimPBr}) introduced
above. In other words, at every point 
$\, ({\bf k}_{1}, \dots, {\bf k}_{d}, \, {\bf U}, \, 
\bm{\theta}_{0}) \, $ 
we can write for a complete set 
$\{ {\bf v}^{(k)}_{[{\bf k}_{1}, \dots, {\bf k}_{d}, {\bf U},
\bm{\theta}_{0}]} ({\bf x}) \}$
of linearly independent quasiperiodic solutions of
(\ref{AnnSolCoordRepr}) the relations:
$$v^{(k)}_{[{\bf k}_{1}, \dots, {\bf k}_{d}, {\bf U},
\bm{\theta}_{0}] \, i} \, ({\bf x}) \,\,\, = \,\,\,
\sum_{\gamma=1}^{m+s} \,\, \gamma^{k}_{\gamma} 
({\bf k}_{1}, \dots, {\bf k}_{d}, \, {\bf U}) \,\, \left.
{\delta I^{\gamma} \over \delta \varphi^{i} ({\bf x})}
\right|_{\Lambda} $$
with some functions
$\, \gamma^{k}_{\gamma} ({\bf k}_{1}, \dots, {\bf k}_{d}, \, {\bf U}) \, $
on $\, \Lambda$.
}

\vspace{0.3cm}

 It should be noted here that the definition given above implies in fact 
that the number of the additional parameters 
$\, (n^{1}, \dots, n^{s}) \, $ on $\, \Lambda \, $ is equal to the
number of annihilators of the bracket (\ref{MultDimPBr}). So, in this 
scheme the additional parameters $\, (n^{1}, \dots, n^{s}) \, $ are
directly connected with the annihilators of the Poisson bracket. 
 
\vspace{0.2cm}

 Like in the one-dimensional case, we can write the following relations
for the time evolution of the densities
$\, P^{\gamma} (\bm{\varphi}, \, \bm{\varphi}_{\bf x}, \, \dots ) \, $:
$$P^{\gamma}_{t} \left( \bm{\varphi}, \, \bm{\varphi}_{\bf x},
\, \dots \right) \,\,\,\,\, = \,\,\,\,\,
Q^{\gamma 1}_{x^{1}} \left( \bm{\varphi}, \,
\bm{\varphi}_{\bf x}, \, \dots \right) \,\, + \,\,
\dots \,\, + \,\, Q^{\gamma d}_{x^{d}} \left( \bm{\varphi}, \,
\bm{\varphi}_{\bf x}, \, \dots \right) $$  

\vspace{0.2cm}

 Let us consider now the modulation equations for a complete
Hamiltonian family $\, \Lambda \, $ equipped with a minimal set of
commuting integrals $\, \{ I^{1}, \dots, I^{m+s} \}$. It is convenient
to choose now the parameters of the slowly modulated solutions of
(\ref{EvInSyst}) in the form 
$$\left( {\bf S} ({\bf X}, T), \, {\bf U} ({\bf X}, T) \right)
\,\,\, = \,\,\, \left( S^{1} ({\bf X}, T), \dots , S^{m} ({\bf X}, T), 
\, U^{1} ({\bf X}, T), \dots , U^{m+s} ({\bf X}, T) \right) \,\,\, , $$
such that the parameters $\, {\bf k}_{q} ({\bf X}, T) \, $ are defined
by the relations $\, {\bf k}_{q} \, = \, {\bf S}_{X^{q}} \, $
$\, $ (${\bf X} \, = \, \epsilon \, {\bf x}$, 
$\, T \, = \, \epsilon \, t$).

 The regular Whitham system can be written now in the following form
\begin{equation}
\label{MultDimConsWhithSyst}
\begin{array}{c}
S^{\alpha}_{T} \,\,\,\,\, = \,\,\,\,\, \omega^{\alpha} \left(
{\bf S}_{X^{1}}, \dots, {\bf S}_{X^{d}}, {\bf U}
\right) \,\,\, , \,\,\,\,\,\,\,\,\,\,\,\,\,\,\, 
\alpha \, = \, 1, \dots, m \,\, ,  \\  \\
U^{\gamma}_{T} \,\,\,\,\, = \,\,\,\,\,
\langle Q^{\gamma 1} \rangle_{X^{1}} \, + \, \dots \, + \,
\langle Q^{\gamma d} \rangle_{X^{d}} \,\,\, , \,\,\,\,\,\,\,\,\,\,
\gamma \, = \, 1, \dots , m + s \,\, ,
\end{array}
\end{equation}
which is equivalent to the system defined by 
(\ref{SOmKRel})-(\ref{KappafOrtRel}) (\cite{MinimalSet}).

 The procedure of averaging of the Poisson bracket (\ref{MultDimPBr})
represents a modification of the Dubrovin - Novikov procedure and can
be formulated in the following way (\cite{JMP2, MinimalSet}):

 Like in the one-dimensional case, let us calculate the pairwise
Poisson brackets of the densities $\, P^{\gamma} ({\bf x})$,
$\, P^{\rho} ({\bf y})$, which can be represented now in the form
$$\{ P^{\gamma} ({\bf x}) \, , \, P^{\rho} ({\bf y}) \} \,\,\, = \,\,\,
\sum_{l_{1},\dots,l_{d}} \, A^{\gamma\rho}_{l_{1} \dots l_{d}}
(\bm{\varphi}, \bm{\varphi}_{\bf x}, \dots ) \,\,\,
\delta^{(l_{1})} (x^{1} - y^{1}) \, \dots \,
\delta^{(l_{d})} (x^{d} - y^{d}) $$
($l_{1}, \dots, l_{d} \geq 0$).

 In the same way, we can write here the relations
$$A^{\gamma\rho}_{0 \dots 0} (\bm{\varphi}, \bm{\varphi}_{\bf x}, \dots )
\,\,\,\,\, \equiv \,\,\,\,\, \partial_{x^{1}} \, Q^{\gamma\rho 1}
(\bm{\varphi}, \bm{\varphi}_{\bf x}, \dots ) \,\, + \,\, \dots \,\, + \,\,
\partial_{x^{d}} \, Q^{\gamma\rho d}
(\bm{\varphi}, \bm{\varphi}_{\bf x}, \dots ) $$
for some functions
$\, (Q^{\gamma\rho 1} (\bm{\varphi}, \bm{\varphi}_{\bf x}, \dots ),
\dots, Q^{\gamma\rho d} (\bm{\varphi}, \bm{\varphi}_{\bf x}, \dots ))$.

 We define the averaged Poisson bracket
$\, \{ \dots , \dots \}_{\rm AV} \, $ on the space of fields
$\, ({\bf S} ({\bf X}), \, {\bf U} ({\bf X})) \, $ by the 
following equalities:
\begin{equation}
\label{AveragedBracket}
\begin{array}{c}
\left\{ S^{\alpha} ({\bf X})  \, , \,
S^{\beta} ({\bf Y}) \right\}_{\rm AV}
\,\,\, =  \,\,\, 0
\,\,\,\,\,\,\,\, , \,\,\,\,\,\,\,\,\,\,\,\,\,\,\,
\alpha, \beta \, = \, 1, \dots , m \, ,  \\  \\
\left\{ S^{\alpha} ({\bf X})  \, , \,
U^{\gamma} ({\bf Y}) \right\}_{\rm AV} \,\,\, = \,\,\,
\omega^{\alpha\gamma}
\left({\bf S}_{X^{1}}, \dots, {\bf S}_{X^{d}},
{\bf U} ({\bf X}) \! \right)
\, \delta ({\bf X} - {\bf Y})  \,\,\,\,\, ,  \\  \\
\left\{ U^{\gamma} ({\bf X})\, , \,
U^{\rho} ({\bf Y}) \right\}_{\rm AV}
\,\,\, = \,\,\, \langle A^{\gamma\rho}_{10\dots0} \rangle
\left({\bf S}_{X^{1}}, \dots, {\bf S}_{X^{d}},
{\bf U} ({\bf X}) \! \right) \,\,
\delta_{X^{1}} ({\bf X} - {\bf Y})
\,\,\, + \,\, \dots \,\, +   \\  \\
+ \,\,\, \langle A^{\gamma\rho}_{0\dots01} \rangle
\left({\bf S}_{X^{1}}, \dots, {\bf S}_{X^{d}},
{\bf U} ({\bf X}) \! \right) \,\,\,
\delta_{X^{d}} ({\bf X} - {\bf Y}) \,\,\, +   \\  \\ 
+ \,\,\, \left[ \langle Q^{\gamma\rho \, p} \rangle
\left({\bf S}_{X^{1}}, \dots, {\bf S}_{X^{d}},
{\bf U} ({\bf X}) \! \right)
\right]_{X^{p}} \,\,\, \delta ({\bf X} - {\bf Y})
\,\,\,\,\,\,\,\, , \,\,\,\,\,\,\,\,\,\,\,\,\,\,\,
\gamma, \rho \, = \, 1, \dots , m + s
\end{array}
\end{equation}

 System (\ref{MultDimConsWhithSyst}) can be written now as a 
Hamiltonian system with the bracket (\ref{AveragedBracket}) and the
Hamiltonian functional
$$H_{av} \,\,\, = \,\,\, \int \, \langle P_{H} \rangle \,
\left( {\bf S}_{X^{1}}, \dots, {\bf S}_{X^{d}},
\, {\bf U} ({\bf X}) \right) \,\, d^{d} X $$

 The detailed consideration and justification of the above procedure
for a complete Hamiltonian family $\, \Lambda \, $ equipped with a
minimal set of commuting integrals can be found in \cite{MinimalSet}.
Let us say, that the same procedure was considered also under some 
other requirements on the family of $m$-phase solutions of 
(\ref{EvInSyst}) in \cite{JMP2}.

 Let us formulate here also a theorem claiming the invariance of the
procedure of the bracket averaging.

\vspace{0.3cm}

\noindent
{\bf Theorem 1.1} (\cite{MinimalSet}).

{\it Let a family $\, \Lambda \, $ represent a complete Hamiltonian 
family of $m$-phase solutions of system (\ref{EvInSyst}) equipped with 
a minimal set of commuting integrals 
$\, \{ I^{1}, \dots , I^{m+s} \}$. Let the set
$\, \{ I^{\prime 1}, \dots , I^{\prime m+s} \} \, $
represent another minimal set of commuting integrals for the
family $\, \Lambda , \, $ satisfying all the requirements of 
Definition 1.3.

 Then the Poisson brackets (\ref{AveragedBracket}), obtained with 
the aid of the sets $\, \{ I^{1}, \dots , I^{m+s} \} \, $ 
and   \linebreak
$\, \{ I^{\prime 1}, \dots , I^{\prime m+s} \} , \, $
coincide with each other.
}

\vspace{0.3cm}

 In other words, under the requirements of Theorem 1.1 we can claim,
that the expressions (\ref{AveragedBracket}), obtained with
the aid of the sets $\, \{ I^{1}, \dots , I^{m+s} \} \, $ and
$\, \{ I^{\prime 1}, \dots , I^{\prime m+s} \} , \, $
transform into each other under the coordinate transformation
\begin{equation}
\label{SUUprimeTrans}
\left( S^{1} ({\bf X}), \dots , S^{m} ({\bf X}), \,
U^{1} ({\bf X}), \dots , U^{m+s} ({\bf X}) \right)
\quad \rightarrow \quad 
\left( S^{1} ({\bf X}), \dots , S^{m} ({\bf X}), \,
U^{\prime 1} ({\bf X}), \dots , U^{\prime m+s} ({\bf X}) \right)
\,\,\, , 
\end{equation}
where $\, (U^{1}, \dots , U^{m+s}) \, $ and
$\, (U^{\prime 1}, \dots , U^{\prime m+s}) \, $ are the parameters
on the family $\, \Lambda , \, $ corresponding to the sets
$\, \{ I^{1}, \dots , I^{m+s} \} \, $ and
$\, \{ I^{\prime 1}, \dots , I^{\prime m+s} \} , \, $
respectively.

\vspace{0.3cm}

 The main purpose of this article is to study the canonical forms
of the brackets (\ref{AveragedBracket}) so we could have an analog 
of the Darboux Theorem for the averaged Poisson brackets in the 
multi-dimensional case. 

 Let us say, however, that we will be interested here just in the 
special coordinate transformations, preserving the ``physical''
meaning of the fields 
$\, ( S^{1} ({\bf X}), \dots , S^{m} ({\bf X}) ) \, $ and
$\, ( U^{1} ({\bf X}), \dots , U^{m+s} ({\bf X}) ) \, $. Thus,
we will always keep here the variables
$\, ( S^{1} ({\bf X}), \dots , S^{m} ({\bf X}) ) , \, $
representing the ``phase'' functions, as the first part of canonical 
variables for the bracket (\ref{AveragedBracket}). So, the
transformations we consider here will have in fact the form
(\ref{SUUprimeTrans}) written above. Besides that, we will always
assume here that the variables
$\, ( U^{1} ({\bf X}), \dots , U^{m+s} ({\bf X}) ) \, $
represent some densities of ``Hydrodynamic Type'', which means in 
fact that transformations 
$\,\, {\bf U} ({\bf X}) \,\, \rightarrow \,\, 
{\bf U}^{\prime} ({\bf X}) \,\, $ should have the ``Hydrodynamic''
form
\begin{equation}
\label{UprimegammaTrans}
U^{\prime \gamma} ({\bf X}) \,\,\, = \,\,\,
U^{\prime \gamma} \left( {\bf S}_{X^{1}}, \dots , {\bf S}_{X^{d}},
\, {\bf U} ({\bf X}) \right) 
\end{equation}

 As we will see in Chapter 2, any bracket (\ref{AveragedBracket})
with the additional condition (\ref{RankProperty}) can be transformed
to the standard canonical form by a transformation 
(\ref{SUUprimeTrans}) - (\ref{UprimegammaTrans}) in the special
case $\, s \, = \, 0 \, . \, $  This case corresponds in fact to the
absence of annihilators of the bracket (\ref{MultDimPBr}) on the space
of the quasiperiodic functions and allows always the construction
of the second part of canonical variables
$\, ( Q_{1} ({\bf X}), \dots , Q_{m} ({\bf X}) ) \, , \, $
conjugated to the variables
$\, ( S^{1} ({\bf X}), \dots , S^{m} ({\bf X}) ) \, $.

 In Chapter 3 we consider more complicated case of the presence
of additional parameters   \linebreak
$\, (n^{1}, \dots , n^{s}) \, $ connected
with the presence of annihilators of the bracket (\ref{MultDimPBr}).
As we will see, the situation is more complicated in this case.
We suggest here a generalization of the canonical form for the
bracket (\ref{AveragedBracket}) which represents the sum of the
standard (``action - angle'') part and an independent Poisson
bracket for some additional variables
$\, ( {\bar N}^{1} ({\bf X}) , \dots , 
{\bar N}^{s} ({\bf X}) ) \, . \, $  As we will show, the averaged
bracket (\ref{AveragedBracket}) can be transformed into the
``pseudo-canonical'' form by a coordinate transformation
(\ref{SUUprimeTrans}) - (\ref{UprimegammaTrans}) under some 
additional (``physical'') requirements on the initial bracket 
(\ref{MultDimPBr}). We have to say, however, that an abstract
Poisson bracket (\ref{AveragedBracket}) can not in general be
written in the pseudo-canonical form after a coordinate
transformation (\ref{SUUprimeTrans}) - (\ref{UprimegammaTrans}) 
which is demonstrated by a special example at the end of Chapter 3.

\section{The Canonical form of the averaged bracket.}
\setcounter{equation}{0}

 First, let us introduce here special coordinates for the bracket
(\ref{AveragedBracket}) which will give a basis for it's further
consideration. Everywhere below we will assume that the bracket
(\ref{AveragedBracket}) represents the averaging of the bracket
(\ref{MultDimPBr}) on a complete Hamiltonian family of $m$-phase 
solutions of system (\ref{EvInSyst}) equipped with a minimal set of 
commuting integrals.

\vspace{0.5cm}

 Let us consider an $(m d + m + s)$-dimensional manifold with 
coordinates $\, ({\bf k}_{1}, \dots, {\bf k}_{d}, \, {\bf U}) \, $
and the $(m + s)$-dimensional submanifolds given by the relations
$\, ({\bf k}_{1}, \dots, {\bf k}_{d}) = {\rm const}$. Consider the 
vector fields
$${\vec \xi}_{(\alpha)} \,\, = \,\,
\left( \, \omega^{\alpha \, 1} 
({\bf k}_{1}, \dots, {\bf k}_{d}, \,  {\bf U}) , \dots,
\, \omega^{\alpha \, m + s} 
({\bf k}_{1}, \dots, {\bf k}_{d}, \,  {\bf U}) \, \right)^{t} $$
on the submanifolds 
$\, ({\bf k}_{1}, \dots, {\bf k}_{d}) = {\rm const}$.

 Using the Jacobi identities
$$\left\{ \left\{ U^{\gamma} ({\bf X}) \, , \, S^{\alpha} ({\bf Y})
\right\} \, , \, S^{\beta} ({\bf Z}) \right\} \,\, - \,\,
\left\{ \left\{ U^{\gamma} ({\bf X}) \, , \, S^{\beta} ({\bf Z})
\right\} \, , \, S^{\alpha} ({\bf Y}) \right\}
\,\,\, \equiv \,\,\, 0 $$
for the bracket (\ref{AveragedBracket}), we easily get the 
following relations
$$\left[ \, {\vec \xi}_{(\alpha)} \, , \, {\vec \xi}_{(\beta)} \,
\right] \,\,\, \equiv \,\,\, 0 \,\,\,\,\, , \,\,\,\,\,\,\,\,
\alpha, \beta = 1, \dots, m $$
for the commutators of the vectors fields 
$\, {\vec \xi}_{(\alpha)} \, $
on the submanifolds 
$\, ({\bf k}_{1}, \dots, {\bf k}_{d}) = {\rm const}$.

  According to relations (\ref{RankProperty}) we can state also that 
the set of vector fields $\{ {\vec \xi}_{(\alpha)} \}$ is linearly 
independent at every point.

 We can claim then that on every submanifold
$\, ({\bf k}_{1}, \dots, {\bf k}_{d}) = {\rm const} \, $
there exists a locally invertible change of coordinates
\begin{multline*}
\left( U^{1}, \dots, U^{m+s} \right)
\,\,\, \rightarrow   \\
\rightarrow \,
\left( {\hat Q}_{1}
({\bf k}_{1}, \dots, {\bf k}_{d}, {\bf U}), \dots, 
{\hat Q}_{m}({\bf k}_{1}, \dots, {\bf k}_{d}, {\bf U}), \,
{\hat N}^{1}({\bf k}_{1}, \dots, {\bf k}_{d}, {\bf U}), \dots, 
{\hat N}^{s}({\bf k}_{1}, \dots, {\bf k}_{d}, {\bf U}) \right) 
\end{multline*}
depending smoothly on the parameters 
$\, ({\bf k}_{1}, \dots, {\bf k}_{d})$,
which leads to the following coordinate representation
$${\vec \xi}_{(1)} \, = \, ( 1, 0, \dots, 0 )^{t} \,\, , \,\,\,\,\,
\dots \,\,\,\,\, , \,\,\,\,\, {\vec \xi}_{(m)} \, = \,
( 0, \dots, 0, 1, 0, \dots, 0  )^{t} $$
of the vector fields
$\, {\vec \xi}_{(\alpha)} \, $ on these submanifolds.

 It is not difficult to see then that for the functionals
$${\tilde Q}_{\alpha} ({\bf X}) \,\, = \,\, 
{\hat Q}_{\alpha} \left( {\bf S}_{X^{1}}, \dots, {\bf S}_{X^{d}},
\, {\bf U} ({\bf X}) \right) \,\,\,\,\, , \,\,\,\,\,\,\,\,\,\,
{\tilde N}^{l} ({\bf X}) \,\, = \,\,
{\hat N}^{l} \left( {\bf S}_{X^{1}}, \dots, {\bf S}_{X^{d}},
\, {\bf U} ({\bf X}) \right) $$
we get immediately the following relations
$$\left\{ S^{\alpha} ({\bf X}) \, , \, 
{\tilde Q}_{\beta} ({\bf Y}) \right\}
\,\, = \,\, \delta^{\alpha}_{\beta} \,\,
\delta ({\bf X} - {\bf Y}) \,\,\,\,\, , \,\,\,\,\,\,\,\,\,\,
\left\{ S^{\alpha} ({\bf X}) \, , \, 
{\tilde N}^{l}  ({\bf Y}) \right\} \,\, = \,\, 0 $$

 The pairwise Poisson brackets of the functionals
$\, {\tilde Q}_{\alpha} ({\bf X})$, $\, {\tilde N}^{l} ({\bf X})$
have a local translationally invariant form which 
we can write in general as
$$\left\{ {\tilde Q}_{\alpha} ({\bf X}) \, , \, 
{\tilde Q}_{\beta} ({\bf Y}) \right\}
\,\, = \,\, J_{\alpha\beta} \, ({\bf X}, \, {\bf Y}) 
\,\,\,\,\, , \,\,\,\,\,\,\,\,
\left\{ {\tilde Q}_{\alpha} ({\bf X}) \, , \, 
{\tilde N}^{l}  ({\bf Y}) \right\}
\,\, = \,\, J_{\alpha}^{l} \, ({\bf X}, \, {\bf Y}) $$
$$\left\{  {\tilde N}^{l}  ({\bf X}) \, , \, 
{\tilde N}^{q}  ({\bf Y}) \right\}
\,\, = \,\, J^{lq} \, ({\bf X}, \, {\bf Y}) $$

 Using now the Jacobi identities
$$\left\{ \left\{ {\tilde Q}_{\alpha} ({\bf X}) ,  
{\tilde Q}_{\beta} ({\bf Y}) \right\} , 
\, S^{\gamma} ({\bf Z}) \right\} \,\, + \,\,
c.p. \,\, \equiv \,\, 0 \,\,\, , \,\,\,\,\,
\left\{ \left\{ {\tilde Q}_{\alpha} ({\bf X}) \, , \, 
{\tilde N}^{l} ({\bf Y}) \right\} , 
\, S^{\gamma} ({\bf Z}) \right\} \,\, + \,\,
c.p. \,\, \equiv \,\, 0 $$
$$\left\{ \left\{ {\tilde N}^{l} ({\bf X}) , {\tilde N}^{q} ({\bf Y})
\right\} , \, S^{\gamma} ({\bf Z}) \right\} \,\, + \,\,
c.p. \,\, \equiv \,\, 0 $$
we obtain also the following relations
$${\delta J_{\alpha\beta} \, ({\bf X}, \, {\bf Y}) \over  
\delta {\tilde Q}_{\gamma} ({\bf Z})} \, \equiv \, 0
\,\,\,\,\, , \,\,\,\,\,\,\,\,
{\delta J_{\alpha}^{l} \, ({\bf X}, \, {\bf Y}) \over
\delta {\tilde Q}_{\gamma} ({\bf Z})} \, \equiv \, 0
\,\,\,\,\, , \,\,\,\,\,\,\,\,
{\delta J^{lq} \, ({\bf X}, \, {\bf Y}) \over 
\delta {\tilde Q}_{\gamma} ({\bf Z})} \, \equiv \, 0
\,\,\,\,\,\,\,\, , \,\,\,\,\,\,\,\,\,\, \gamma = 1, \dots, m $$
for the distributions
$\, J_{\alpha\beta} \, ({\bf X}, \, {\bf Y}) \, , \, $ 
$\, J_{\alpha}^{l} \, ({\bf X}, \, {\bf Y}) \, , \, $ 
$\, J^{lq} \, ({\bf X}, \, {\bf Y}) \, $.

\vspace{0.3cm}

 Finally, we can write the Poisson bracket (\ref{AveragedBracket})
in coordinates $\, {\bf S} ({\bf X})$, 
$\, {\tilde Q}_{\alpha} ({\bf X})$, 
$\, {\tilde N}^{l} ({\bf X}) \, $
in the following general form
$$\left\{ S^{\alpha} ({\bf X}) \, , \, S^{\beta} ({\bf Y}) \right\}
\,\,\, = \,\,\, 0 $$
$$\left\{ S^{\alpha} ({\bf X}) \, , \,
{\tilde Q}_{\beta} ({\bf Y}) \right\}
\,\,\, = \,\,\, \delta^{\alpha}_{\beta} \,\,
\delta ({\bf X} - {\bf Y}) \,\,\,\,\, , \,\,\,\,\,\,\,\,\,\,
\left\{ S^{\alpha} ({\bf X}) \, , \,
{\tilde N}^{l}  ({\bf Y}) \right\} \,\,\,\ = \,\,\, 0 $$ 
\begin{multline*}
\left\{ {\tilde Q}_{\alpha} ({\bf X}) \, , \,
{\tilde Q}_{\beta} ({\bf Y}) \right\}
\,\,\,\,\, = \,\,\,\,\, \Omega_{\alpha\beta}^{p} 
( {\bf S}_{\bf X}, {\tilde {\bf N}} ) \,\,\, 
\delta_{X^{p}} ({\bf X} - {\bf Y}) \,\,\, +  \\
+ \,\,\, \Gamma_{\alpha\beta\gamma}^{pq}
( {\bf S}_{\bf X}, {\tilde {\bf N}} ) \,\,
S^{\gamma}_{X^{p}X^{q}} \,\,\, \delta ({\bf X} - {\bf Y}) 
\,\,\, + \,\,\, \Pi_{\alpha\beta r}^{p} 
( {\bf S}_{\bf X}, {\tilde {\bf N}} ) \,\,
{\tilde N}^{r}_{X^{p}} \,\,\, \delta ({\bf X} - {\bf Y}) 
\end{multline*}
$(\Gamma_{\alpha\beta\gamma}^{pq} \, \equiv \, 
\Gamma_{\alpha\beta\gamma}^{qp})$,
\begin{multline*}
\left\{ {\tilde Q}_{\alpha} ({\bf X}) \, , \,
{\tilde N}^{l}  ({\bf Y}) \right\}
\,\,\,\,\, = \,\,\,\,\, A_{\alpha}^{l,p}
( {\bf S}_{\bf X}, {\tilde {\bf N}} ) \,\,\,
\delta_{X^{p}} ({\bf X} - {\bf Y}) \,\,\, +  \\
+ \,\,\, B_{\alpha\gamma}^{l,pq}
( {\bf S}_{\bf X}, {\tilde {\bf N}} ) \,\, 
S^{\gamma}_{X^{p}X^{q}} \,\,\, 
\delta ({\bf X} - {\bf Y}) \,\,\, + \,\,\,
C_{\alpha r}^{l,p} ( {\bf S}_{\bf X}, {\tilde {\bf N}} ) \,\,
{\tilde N}^{r}_{X^{p}} \,\,\, \delta ({\bf X} - {\bf Y})
\end{multline*}
$(B_{\alpha\gamma}^{l,pq} \, \equiv \, B_{\alpha\gamma}^{l,qp})$,
\begin{multline*}
\left\{  {\tilde N}^{l}  ({\bf X}) \, , \,
{\tilde N}^{k}  ({\bf Y}) \right\} 
\,\,\,\,\, = \,\,\,\,\, g^{lk,p}
( {\bf S}_{\bf X}, {\tilde {\bf N}} ) \,\,\,
\delta_{X^{p}} ({\bf X} - {\bf Y}) \,\,\, +  \\
+ \,\,\, b_{r}^{lk,p} ( {\bf S}_{\bf X}, {\tilde {\bf N}} ) \,\,
{\tilde N}^{r}_{X^{p}} \,\,\, \delta ({\bf X} - {\bf Y}) 
\,\,\, + \,\,\,
M_{\gamma}^{lk,pq} ( {\bf S}_{\bf X}, {\tilde {\bf N}} ) \,\,
S^{\gamma}_{X^{p}X^{q}} \,\,\, \delta ({\bf X} - {\bf Y}) 
\end{multline*}
$(M_{\gamma}^{lk,pq} \, \equiv \, M_{\gamma}^{lk,qp})$.

\vspace{0.3cm}

 Easy to see that, according to their definition, the variables
$\, {\tilde Q}_{\alpha} ({\bf X}) \, $ and 
$\, {\tilde N}^{l} ({\bf X}) \, $
are defined modulo the transformations
$${\tilde Q}_{\alpha} ({\bf X}) \,\,\,\,\, \rightarrow \,\,\,\,\,
{\tilde Q}_{\alpha} ({\bf X}) \,\, + \,\, {\tilde f}_{\alpha}
\left( {\bf S}_{\bf X}, \, {\tilde {\bf N}} ({\bf X}) \right) 
\,\,\,\,\, , \,\,\,\,\,\,\,\,\,\, 
{\tilde N}^{l} ({\bf X}) \,\,\,\,\, \rightarrow \,\,\,\,\,
{\tilde N}^{\prime l} \left( {\bf S}_{\bf X}, \,
{\tilde {\bf N}} ({\bf X}) \right) $$
where
$${\rm det} \,\, \left| \left|
{\partial {\tilde N}^{\prime l} \over \partial {\tilde N}^{k}}
\right| \right| \,\,\, \neq \,\,\, 0 $$

\vspace{0.2cm}

 It is not difficult to see also that the values 
$\, \{  {\tilde N}^{l}  ({\bf X}) \, , \, 
{\tilde N}^{k}  ({\bf Y}) \} \, $ define a Poisson bracket on the
space of fields $\, {\tilde {\bf N}} ({\bf X}) \, $ at any fixed values
of $\, ( S^{1} ({\bf X}), \dots , S^{m} ({\bf X}))$.

\vspace{0.3cm}

 We will consider in this chapter an important case when the number
of annihilators of the bracket (\ref{MultDimPBr}) and the number of
additional parameters $\, ( n^{1}, \dots , n^{s} ) \, $ on 
$\, \Lambda \, $ are equal to zero. As we will see, the investigation
of the canonical form of the bracket (\ref{AveragedBracket}) represents
in this case a special interest.

\vspace{0.2cm}

 Let us write down the averaged bracket
(\ref{AveragedBracket}) in coordinates
$\, (S^{\alpha} ({\bf X}) \, , \, {\tilde Q}_{\alpha} ({\bf X})) , \, $
such that we will have
$$\quad \quad \quad \quad  \left\{ S^{\alpha} ({\bf X}) \, , \, 
S^{\beta} ({\bf Y}) \right\}
\,\, = \,\, 0 \,\,\,\,\, , \,\,\,\,\,\,\,\,
\left\{ S^{\alpha} ({\bf X}) \, , \,
{\tilde Q}_{\beta} ({\bf Y}) \right\}
\,\, = \,\, \delta^{\alpha}_{\beta} \,\,
\delta ({\bf X} - {\bf Y}) \,\,\,\,\, ,  \quad \quad \quad \quad  $$
\begin{equation}
\label{AdmBr}
\left\{ {\tilde Q}_{\alpha} ({\bf X}) \, , \,
{\tilde Q}_{\beta}  ({\bf Y}) \right\}
\,\,\, = \,\,\, J_{\alpha\beta} \left[ {\bf S} \right]
({\bf X}, \, {\bf Y}) \,\,\, = \,\,\, \Omega_{\alpha\beta}^{p}
\left( {\bf S}_{\bf X} \right) \,\,
\delta_{X^{p}} ({\bf X} - {\bf Y}) \,\, + \,\, 
\Gamma_{\alpha\beta\gamma}^{pq}
\left( {\bf S}_{\bf X} \right) \,
S^{\gamma}_{X^{p}X^{q}} \,\, \delta ({\bf X} - {\bf Y}) 
\end{equation}

 The Jacobi identities
$$\left\{ \left\{ {\tilde Q}_{\alpha} ({\bf X}) \, , \,
{\tilde Q}_{\beta} ({\bf Y}) \right\} \, , \,   
{\tilde Q}_{\gamma} ({\bf Z}) \right\} \,\,\,\,\, + \,\,\,\,\,
c.p. \,\,\,\,\, \equiv \,\,\,\,\, 0 $$
give now the following relations
\begin{equation}
\label{ClosedJRelations}
{ \delta J_{\alpha\beta} [{\bf S}] ({\bf X}, {\bf Y})
\over \delta S^{\gamma} ({\bf Z})} \,\,\, + \,\,\,
{ \delta J_{\beta\gamma} [{\bf S}] ({\bf Y}, {\bf Z})
\over \delta S^{\alpha} ({\bf X})} \,\,\, + \,\,\,
{ \delta J_{\gamma\alpha} [{\bf S}] ({\bf Z}, {\bf X})   
\over \delta S^{\beta} ({\bf Y})} \,\,\,\,\, \equiv \,\,\,\,\, 0
\end{equation}
for the functionals
$\, J_{\alpha\beta} [{\bf S}] ({\bf X}, {\bf Y})$,
which mean the closeness of the two-form 
\begin{equation}
\label{OprFormJ}
\int \, J_{\alpha\beta} [{\bf S}] \, ({\bf X}, {\bf Y}) \,\,\,
\delta S^{\alpha} ({\bf X})  \wedge  
\delta S^{\beta} ({\bf Y}) \,\,\, d^{d} X \, d^{d} Y
\end{equation}
on the space of fields $\, (S^{1} ({\bf X}), \dots, S^{m} ({\bf X}))$. 

\vspace{0.2cm}

 According to the terminology of S.P. Novikov (\cite{NovikovMultVal}),
the brackets of the form (\ref{AdmBr}) represent 
``variationally admissible'' Poisson brackets, connected with the 
Lagrangian representation for the corresponding Hamiltonian systems. 
As was shown in \cite{NovikovMultVal}, the variationally admissible 
Poisson brackets lead in general to a nontrivial Lagrangian 
representation of the Hamiltonian systems where the Lagrangian represents 
in fact a closed 1-form on the functional space. As can be also shown, the 
variationally admissible Hamiltonian structures have in general rather 
nontrivial topological invariants connected with the topology of the 
functional space (\cite{NovikovMultVal}).

 Let us say that in general case we can admit that the variables 
$\, S^{\alpha} ({\bf X}) \, $ represent ``unobservable'' quantities,
such that only their spatial and time derivatives can appear in all
``physically measurable'' values. As a corollary, we can admit also,
that only the spatial and time derivatives of the functions
$\, (S^{1} ({\bf X}), \dots, S^{m} ({\bf X})) , \, $ but not the 
functions $\, {\bf S} ({\bf X}) \, $ themselves, are in general
uniquely defined for solutions of the corresponding Hamiltonian systems.
Certainly, the most important class of solutions of this kind is
represented by solutions containing $\, (d-1)$-dimensional 
singularities (``vortices'') in $\, {\bf X}$-space, where the functions 
$\, {\bf S} ({\bf X}) \, $ are not defined, while the increments of
the functions $\, S^{\alpha} ({\bf X}) \, $ along any closed
1-dimensional contour surrounding the vortex are different from zero.

 According to the circumstance mentioned above we will separately
consider here the values having immediate ``physical'' meaning.
As examples of the variables of this kind we can mention here the values 
$\, U^{\gamma} ({\bf X})$, $\, {\tilde Q}_{\alpha} ({\bf X}) \, $ or
the derivatives $\, S^{\alpha}_{X^{q}}$.

 Let us formulate now the theorem about the canonical form of the
bracket (\ref{AdmBr}).

\vspace{0.3cm}

\noindent
{\bf Theorem 2.1.}

{\it For every bracket (\ref{AdmBr}) there exists locally a change 
of coordinates
\begin{equation}
\label{ChangeCoordQ}
Q_{\alpha} ({\bf X}) \,\,\,\,\, = \,\,\,\,\,
{\tilde Q}_{\alpha} ({\bf X}) \,\,\, + \,\,\,
{\tilde q}_{\alpha} \left( {\bf S}_{X^{1}}, \dots, 
{\bf S}_{X^{d}} \right) 
\end{equation}
which transforms bracket (\ref{AdmBr}) into the form
\begin{equation}
\label{StandPBrCanForm}
\left\{ S^{\alpha} ({\bf X}) \, , \, S^{\beta} ({\bf Y}) \right\}
\, = \, 0 \,\,\, , \,\,\,\,\,\,\,
\left\{ S^{\alpha} ({\bf X}) \, , \, Q_{\beta} ({\bf Y}) \right\}
\, = \, \delta^{\alpha}_{\beta} \,\,
\delta ({\bf X} - {\bf Y}) \,\,\, , \,\,\,\,\,\,\,
\left\{ Q_{\alpha} ({\bf X}) \, , \, Q_{\beta}  ({\bf Y}) \right\}
\, = \, 0 
\end{equation}

}

\vspace{0.3cm}

 According to Theorem 2.1 we can claim that for every bracket 
(\ref{AdmBr}) there exist canonical variables 
$\, ( Q_{1} ({\bf X}), \dots , Q_{m} ({\bf X}) ) \, $
conjugated to the variables  
$\, ( S^{1} ({\bf X}), \dots , S^{m} ({\bf X}) ) , \, $
which can be chosen among the ``physically observable'' fields.

\vspace{0.3cm}

 It can be easily seen that Theorem 2.1 permits us to state
also the following theorem about the bracket (\ref{AveragedBracket}):

\vspace{0.3cm}

\noindent
{\bf Theorem 2.1$^{\prime}$.}

{\it Let the relations (\ref{AveragedBracket}) represent a Poisson 
bracket on the space of $2 m$ fields
$$(S^{1} ({\bf X}), \dots, S^{m} ({\bf X}), \,
U^{1} ({\bf X}), \dots, U^{m} ({\bf X})) $$
satisfying the conditions
$${\rm rk} \,\, \left| \left| \, \omega^{\alpha \gamma} \,
({\bf k}_{1}, \dots, {\bf k}_{d}, \, {\bf U}) \, \right| \right| 
\,\,\, = \,\,\, m $$

 Then there exists locally an invertible change of coordinates
\begin{equation}
\label{QSXTrans1}
\left( S^{1}({\bf X}), \dots, S^{m}({\bf X}), \, U^{1}({\bf X}),
\dots, U^{m}({\bf X}) \right) \,\,\,\,\, \rightarrow  \,\,\,\,\,
\left( S^{1}({\bf X}), \dots, S^{m}({\bf X}),
\, Q_{1}({\bf X}), \dots, Q_{m}({\bf X}) \right)
\end{equation}
where
\begin{equation}
\label{QSXTrans2}
Q_{\alpha} \left( {\bf X} \right) \,\,\, = \,\,\,
Q_{\alpha} \left( {\bf S}_{X^{1}}, \dots, {\bf S}_{X^{d}}, \,
{\bf U} ({\bf X}) \right) \,\,\, ,
\end{equation}
such that bracket (\ref{AveragedBracket}) has in the
coordinates $\, ({\bf S} ({\bf X}) \, , \, {\bf Q} ({\bf X}))$
the non-degenerate canonical form:
$$ \left\{ S^{\alpha} ({\bf X}) \, , \, S^{\beta} ({\bf Y}) \right\}
\,\, = \,\, 0 \,\,\, , \,\,\,\,\,
\left\{ S^{\alpha} ({\bf X}) \, , \, Q_{\beta} ({\bf Y}) \right\}
\,\, = \,\, \delta^{\alpha}_{\beta} \,\,
\delta ({\bf X} - {\bf Y}) \,\,\, , \,\,\,\,\,
\left\{ Q_{\alpha} ({\bf X}) \, , \, Q_{\beta}  ({\bf Y}) \right\}
\,\, = \,\, 0 $$

}

\vspace{0.3cm}

 As we said already, Theorem 2.1$^{\prime}$ corresponds to the 
special case, when the number of annihilators of the bracket 
(\ref{MultDimPBr}) and the number of the additional parameters 
$\, (n^{1}, \dots, n^{s})$ on $\Lambda$ are equal to zero.

\vspace{0.1cm}

 Theorem 2.1$^{\prime}$ was first formulated in \cite{JMP2} with
a brief idea of the proof. We will give in this chapter a detailed 
proof of Theorem 2.1 which will imply also Theorem 2.1$^{\prime}$
as a corollary.

\vspace{0.2cm}

 It seems that the most compact proof of Theorem 2.1 can be given
by combining of the geometric and pure computational methods. 
We will assume here for simplicity that all the coefficients of the
bracket (\ref{AdmBr}) represent global smooth functions of their
arguments. Let us note that the proof can be easily modified also for
a local smooth dependence of the coefficients of (\ref{AdmBr}) on the
functions $\, {\bf S}_{\bf X} $. For the proof we will need to prove 
first two following lemmas:

\vspace{0.3cm}

\noindent
{\bf Lemma 2.1.}

{\it Any divergence-free vector field $\, \xi^{r} ({\bf X}) \, $
having the form
$$\xi^{r} ({\bf X}) \,\,\, = \,\,\, F^{qp, r}_{\alpha} \left(
{\bf S}_{\bf X} \right) \, S^{\alpha}_{X^{q}X^{p}}  $$
($F^{qp, r}_{\alpha} \, \equiv \, F^{pq, r}_{\alpha}$), can be 
locally represented in the form  
$$\xi^{r} ({\bf X}) \,\,\,\,\, \equiv \,\,\,\,\, \sum_{s \neq r} \, 
\Big[ f^{sr} \left( {\bf S}_{\bf X} \right) \Big]_{X^{s}} \,\,\, , $$
where 
$\, f^{sr} ({\bf S}_{\bf X}) \, \equiv \, - f^{rs} ({\bf S}_{\bf X})$.

}

\vspace{0.3cm}

Proof.  

From the conditions
$$\sum_{r} \, \Big( F^{qp, r}_{\alpha} \left( {\bf S}_{\bf X} \right) 
\, S^{\alpha}_{X^{q}X^{p}} \Big)_{X^{r}} \,\,\,\,\, \equiv 
\,\,\,\,\, 0 $$
we can get, in particular, the following relations
\begin{equation}
\label{FRelations2}
F^{qq, q}_{\alpha} \left( {\bf S}_{\bf X} \right) 
\,\,\, \equiv \,\,\, 0 \,\,\,\,\, , \,\,\,\,\,\,\,\,
F^{qq, r}_{\alpha} \left( {\bf S}_{\bf X} \right)
\,\,\, \equiv \,\,\, - \, 2 \, F^{qr, q}_{\alpha} 
\left( {\bf S}_{\bf X} \right) \,\,\, , \,\,\,\,\,\,\,\, q \neq r
\end{equation}
\begin{equation}
\label{FRelations3}
{\partial F^{qq, r}_{\alpha} \over \partial S^{\beta}_{X^{q}}}
\,\,\,\,\, \equiv \,\,\,\,\, - \,\, 2 \,\, 
{\partial F^{qr, q}_{\beta} \over \partial S^{\alpha}_{X^{q}}}
\,\,\,\,\, \equiv \,\,\,\,\,
{\partial F^{qq, r}_{\beta} \over \partial S^{\alpha}_{X^{q}}}
\,\,\, , \,\,\,\,\,\,\,\,\,\, q \neq r
\end{equation}
(no summation).
\begin{equation}
\label{FRelations4}
{\partial F^{qq, r}_{\alpha} \over \partial S^{\beta}_{X^{r}}}
\,\,\,\,\, \equiv \,\,\,\,\, - \,\, 
{\partial F^{rr, q}_{\beta} \over \partial S^{\alpha}_{X^{q}}}
\,\,\, , \,\,\,\,\,\,\,\,\,\, q \neq r
\end{equation}
(no summation).

 Form relations (\ref{FRelations3}) and (\ref{FRelations4}) we can
conclude that locally there exist functions 
$\, g^{qr} ({\bf S}_{\bf X}) , \, $ satisfying the relations
\begin{equation}
\label{FgRelation1}
F^{qq, r}_{\alpha} \left( {\bf S}_{\bf X} \right) \,\,\, \equiv \,\,\, 
{\partial g^{qr} \over \partial S^{\alpha}_{X^{q}}}
\,\,\,\,\, , \,\,\,\,\,\,\,\,\,\, g^{qr} \left( {\bf S}_{\bf X} \right)
\,\,\, \equiv \,\,\, - \, g^{rq} \left( {\bf S}_{\bf X} \right)
\end{equation}

 We easily get then also from (\ref{FRelations2}) the relations
\begin{equation}
\label{FgRelation2}
F^{qr, q}_{\alpha} \left( {\bf S}_{\bf X} \right) \,\,\, \equiv \,\,\,
- \, {1 \over 2} \,\, {\partial g^{qr} \over \partial S^{\alpha}_{X^{q}}}
\,\,\, , \,\,\,\,\,\,\,\,\,\, q \neq r
\end{equation} 

 Let us consider now the vector field
$${\tilde \xi}^{r} ({\bf X}) \,\,\,\,\,\,\,\, = \,\,\,\,\,\,\,\,
\xi^{r} ({\bf X}) \,\,\, - \,\,\,  \sum_{q \neq r} \,
\Big[ g^{qr} \left( {\bf S}_{\bf X} \right) \Big]_{X^{q}} 
\,\,\,\,\,\,\,\, = \,\,\,\,\,\,\,\, \xi^{r} ({\bf X}) \,\,\, - \,\,\,  
\sum_{q \neq r} \,\, 
{\partial g^{qr} \over \partial S^{\alpha}_{X^{s}}}
\,\, S^{\alpha}_{X^{s}X^{q}} $$

 Using relations (\ref{FgRelation1}) and (\ref{FgRelation2}) we
conclude now that the field $\, {\tilde \xi}^{r} ({\bf X}) \, $
represents a divergence-free vector field having the form
$${\tilde \xi}^{r} ({\bf X}) \,\,\,\,\, = \,\,\,\,\, \sum \,\,
{\tilde F}^{qp, r}_{\alpha} \left( {\bf S}_{\bf X} \right) \,
S^{\alpha}_{X^{q}X^{p}} $$
where
$${\tilde F}^{qq, r}_{\alpha} \,\,\,\,\, \equiv \,\,\,\,\,
{\tilde F}^{qr, q}_{\alpha} \,\,\,\,\, \equiv \,\,\,\,\, 0 
\,\,\,\,\, , \,\,\,\,\,\,\,\,\,\, 
({\tilde F}^{qp, r}_{\alpha} \,\, \equiv \,\, 
{\tilde F}^{pq, r}_{\alpha}) \,\, . $$

 From the relations
$$\sum_{q \neq p \neq r} \, 
\Big( {\tilde F}^{qp, r}_{\alpha} \left( {\bf S}_{\bf X} \right)
\, S^{\alpha}_{X^{q}X^{p}} \Big)_{X^{r}} \,\,\,\,\, \equiv
\,\,\,\,\, 0 $$
we now get the relations
\begin{equation}
\label{FTildeRel1}
{\tilde F}^{qp, r}_{\alpha} \,\,\, + \,\,\,
{\tilde F}^{rp, q}_{\alpha} \,\,\, + \,\,\,
{\tilde F}^{qr, p}_{\alpha} \,\,\,\,\, \equiv  \,\,\,\,\, 0 
\end{equation}
and
\begin{equation}
\label{FTildeRel2}
{\partial {\tilde F}^{qp, r}_{\alpha} \over 
\partial S^{\beta}_{X^{s}}} \,\,\, + \,\,\,
{\partial {\tilde F}^{qp, s}_{\alpha} \over
\partial S^{\beta}_{X^{r}}} \,\,\, + \,\,\,
{\partial {\tilde F}^{rs, p}_{\beta} \over
\partial S^{\alpha}_{X^{q}}} \,\,\, + \,\,\,
{\partial {\tilde F}^{rs, q}_{\beta} \over
\partial S^{\alpha}_{X^{p}}} \,\,\,\,\, \equiv  \,\,\,\,\, 0
\end{equation}

 In particular, for $\, s \, = \, r \, $ we have:
\begin{equation}
\label{FTildeXrRel}
{\partial {\tilde F}^{qp, r}_{\alpha} \over
\partial S^{\beta}_{X^{r}}} \,\,\,\,\, \equiv  \,\,\,\,\, 0
\end{equation}
(no summation).

 Applying the operator 
$\, \partial^{2} / \partial S^{\beta}_{X^{q}} 
\partial S^{\gamma}_{X^{p}} \, $ to the relations 
(\ref{FTildeRel1}) we easily get the relations:
$${\partial^{2} \, {\tilde F}^{qp, r}_{\alpha} \over
\partial S^{\beta}_{X^{q}} \, \partial S^{\gamma}_{X^{p}}}
\,\,\,\,\, \equiv  \,\,\,\,\, 0 $$
(no summation).

 Using also relations (\ref{FTildeXrRel}) we then get locally the
following representation for $\, {\tilde F}^{qp, r}_{\alpha} \, $:
$${\tilde F}^{qp, r}_{\alpha} \,\,\,\,\, = \,\,\,\,\, a^{qp, r}_{\alpha}
\left( {\bf S}_{X^{1}}, \, \dots , \, {\hat {\bf S}_{X^{q}}}, \,
\dots , \, {\hat {\bf S}_{X^{r}}}, \, \dots , \, {\bf S}_{X^{d}} \right)
\,\,\, +  \,\,\, 
a^{pq, r}_{\alpha} \left( {\bf S}_{X^{1}}, \, \dots , \, 
{\hat {\bf S}_{X^{p}}}, \, \dots , \, {\hat {\bf S}_{X^{r}}}, \, 
\dots , \, {\bf S}_{X^{d}} \right) $$
where the hat over the variable means the absence of this variable
among the arguments of a functions.

 For the functions $\, a^{qp, r}_{\alpha} \, $ we get now the
relations:
\begin{multline*}
a^{qp, r}_{\alpha} \left( {\bf S}_{X^{1}}, \, \dots , \, 
{\hat {\bf S}_{X^{q}}}, \, \dots , \, {\hat {\bf S}_{X^{r}}}, \, 
\dots , \, {\bf S}_{X^{d}} \right) \,\,\, +  \,\,\, a^{pq, r}_{\alpha} 
\left( {\bf S}_{X^{1}}, \, \dots , \, {\hat {\bf S}_{X^{p}}}, \, 
\dots , \, {\hat {\bf S}_{X^{r}}}, \, \dots , \, 
{\bf S}_{X^{d}} \right) \,\,\, +   \\
+ \,\,\, a^{rp, q}_{\alpha} \left( {\bf S}_{X^{1}}, \, \dots , \,
{\hat {\bf S}_{X^{q}}}, \, \dots , \, {\hat {\bf S}_{X^{r}}}, \,
\dots , \, {\bf S}_{X^{d}} \right) \,\,\, +  \,\,\, a^{pr, q}_{\alpha}
\left( {\bf S}_{X^{1}}, \, \dots , \, {\hat {\bf S}_{X^{q}}}, \,
\dots , \, {\hat {\bf S}_{X^{p}}}, \, \dots , \,
{\bf S}_{X^{d}} \right) \,\,\, +   \\
+ \,\,\, a^{qr, p}_{\alpha} \left( {\bf S}_{X^{1}}, \, \dots , \,
{\hat {\bf S}_{X^{q}}}, \, \dots , \, {\hat {\bf S}_{X^{p}}}, \,
\dots , \, {\bf S}_{X^{d}} \right) \,\,\, +  \,\,\,  a^{rq, p}_{\alpha}
\left( {\bf S}_{X^{1}}, \, \dots , \, {\hat {\bf S}_{X^{p}}}, \,
\dots , \, {\hat {\bf S}_{X^{r}}}, \, \dots , \,
{\bf S}_{X^{d}} \right) \,\,\,\,\, \equiv \,\,\,\,\, 0
\end{multline*}

 Applying the operators $\, \partial / \partial S^{\beta}_{X^{q}} \, $
to the relations above, we easily get now the following relations:
\begin{multline}
\label{aSkewSymRel}
a^{pq, r}_{\alpha} \left( {\bf S}_{X^{1}}, \, \dots , \,
{\hat {\bf S}_{X^{p}}}, \, \dots , \, {\hat {\bf S}_{X^{r}}}, \,
\dots , \, {\bf S}_{X^{d}} \right) \,\,\,\,\, =  \,\,\,\,\, 
- \,\, a^{rq, p}_{\alpha} \left( {\bf S}_{X^{1}}, \, \dots , \,
{\hat {\bf S}_{X^{p}}}, \, \dots , \, {\hat {\bf S}_{X^{r}}}, \,
\dots , \, {\bf S}_{X^{d}} \right) \,\,\, +  \\
+ \,\,\,\,\, {\tilde a^{pq, r}_{\alpha}} \left( {\bf S}_{X^{1}}, \, 
\dots , \, {\hat {\bf S}_{X^{q}}}, \, \dots , \, 
{\hat {\bf S}_{X^{p}}}, \, \dots , \, \, {\hat {\bf S}_{X^{r}}}, \,
\dots , \, {\bf S}_{X^{d}} \right) \quad
\end{multline}
 
 Putting now $\, s \, = \, q \, $ in the relations (\ref{FTildeRel2})
we have:
$${\partial {\tilde F}^{qp, r}_{\alpha} \over
\partial S^{\beta}_{X^{q}}} \,\,\, + \,\,\,
{\partial {\tilde F}^{rq, p}_{\beta} \over
\partial S^{\alpha}_{X^{q}}} \,\,\,\,\, \equiv \,\,\,\,\, 0 
\quad \quad \quad {\rm (no \,\,\, summation)} \, , $$
which gives the following relations
$${\partial a^{pq, r}_{\alpha} \over
\partial S^{\beta}_{X^{q}}} \,\,\, + \,\,\,
{\partial a^{rq, p}_{\beta} \over
\partial S^{\alpha}_{X^{q}}} \,\,\,\,\, \equiv \,\,\,\,\, 0
\quad \quad \quad {\rm (no \,\,\, summation)}  $$
for the functions $\, a^{pq, r}_{\alpha}$. From the relations
(\ref{aSkewSymRel}) we get then
$${\partial a^{pq, r}_{\alpha} \over
\partial S^{\beta}_{X^{q}}} \,\,\, - \,\,\,
{\partial a^{pq, r}_{\beta} \over
\partial S^{\alpha}_{X^{q}}} \,\,\,\,\, \equiv \,\,\,\,\, 0 
\quad \quad \quad {\rm (no \,\,\, summation)} \, , $$
which gives locally
$$a^{pq, r}_{\alpha}
\left( {\bf S}_{X^{1}}, \, \dots , \, {\hat {\bf S}_{X^{p}}}, \,
\dots , \, {\hat {\bf S}_{X^{r}}}, \, \dots , \,
{\bf S}_{X^{d}} \right) \,\,\,\,\, \equiv \,\,\,\,\,
{\partial \, h^{p, r} ({\bf S}_{X^{1}}, \, \dots , \, 
{\hat {\bf S}_{X^{p}}}, \, \dots , \, {\hat {\bf S}_{X^{r}}}, \, 
\dots , \, {\bf S}_{X^{d}} ) \over \partial S^{\alpha}_{X^{q}}} $$
for some functions 
$\, h^{p, r} ({\bf S}_{X^{1}}, \, \dots , \,
{\hat {\bf S}_{X^{p}}}, \, \dots , \, {\hat {\bf S}_{X^{r}}}, \,
\dots , \, {\bf S}_{X^{d}} ) \, $ , 
$\,\,\,\,\, h^{p, r} \, \equiv \, - \, h^{r, p}$.

 We get then
$$\sum_{p} \, \Big[ \, 2 \, h^{p, r} \, \Big]_{X^{p}} 
\,\,\,\, = \,\,\,\, \sum_{p} \,\, 2 \,\, 
{\partial h^{p, r} \over \partial S^{\alpha}_{X^{q}}} \,
S^{\alpha}_{X^{q}X^{p}} \,\,\,\, = \,\,\,\, 
2 \, a^{pq, r}_{\alpha} \,
S^{\alpha}_{X^{q}X^{p}} \,\,\,\, = \,\,\,\, 
{\tilde F}^{qp, r}_{\alpha} \,
S^{\alpha}_{X^{q}X^{p}} \,\,\,\, = \,\,\,\, 
{\tilde \xi}^{r} ({\bf X}) $$

 Finally, we obtain
$$\xi^{r} ({\bf X}) \,\,\,\,\, \equiv \,\,\,\,\,
\sum_{p \neq r} \, \Big[ \, g^{pr} \left( {\bf S}_{\bf X} \right) 
\Big]_{X^{p}} \,\,\, + \,\,\, \sum_{p \neq r} \, 
\Big[ \, 2 \, h^{p, r} \left( {\bf S}_{\bf X} \right) \Big]_{X^{p}} $$
which gives the proof of the lemma.

{\hfill Lemma 2.1 is proved.}

\vspace{0.3cm}

\noindent
{\bf Lemma 2.2.}

{\it Let a closed 1-form on the space of functions
$\, {\bf S} ({\bf X}) \, = \,
(S^{1} ({\bf X}), \dots, S^{m} ({\bf X})) \, $ have the form
\begin{equation}
\label{qOneForm}
{\bf q} \,\,\,\,\, = \,\,\,\,\, \int \, q_{\alpha} ({\bf X}) \,\,
\delta S^{\alpha} ({\bf X}) \,\, d^{d} X \,\,\,\,\, \equiv \,\,\,\,\,
\int \, M^{qp}_{\alpha\gamma} \left( {\bf S}_{\bf X} \right) \,
S^{\gamma}_{X^{q}X^{p}} \,\,\, 
\delta S^{\alpha} ({\bf X}) \,\, d^{d} X 
\end{equation}
($M^{qp}_{\alpha\gamma} \, \equiv \, M^{pq}_{\alpha\gamma}$).

 Then the functions 
$\, q_{\alpha} ({\bf X}) \,\, \equiv \,\, M^{qp}_{\alpha\gamma}
({\bf S}_{\bf X}) \, S^{\gamma}_{X^{q}X^{p}} \, $
can be locally represented as
$$q_{\alpha} ({\bf X}) \,\,\,\,\, \equiv \,\,\,\,\,
{\delta \over \delta S^{\alpha} ({\bf X})} \, \int \,
h \left( {\bf S}_{\bf W} \right) \,\, d^{d} W $$
with some smooth function $\, h ({\bf S}_{\bf W})$. 

}

\vspace{0.3cm}

Proof.

Using the standard ``homotopy operator'' (see e.g. \cite{Olver}),
based on the mapping:
$${\hat F} \,\, : \quad 
\left[ 0 , 1 \right] \times \left\{ {\bf S} ({\bf X}) \right\}
\quad \rightarrow \quad \left\{ {\bf S} ({\bf X}) \right\} \quad ,
\quad \quad \left( \lambda \, , \, {\bf S} ({\bf X}) \right) \quad
\rightarrow \quad \lambda \, {\bf S} ({\bf X}) \,\,\, ,
\,\,\,\,\, \lambda \, \in \, \left[ 0 , 1 \right] \,\,\, , $$
we can claim that the coefficients $ \, q_{\alpha} ({\bf X}) \, $
can be written in the form
$$M^{qp}_{\alpha\gamma} ({\bf S}_{\bf X}) \, S^{\gamma}_{X^{q}X^{p}} 
\,\,\,\,\, \equiv \,\,\,\,\, {\delta \over \delta S^{\alpha} ({\bf X})} 
\, \int \, S^{\rho} ({\bf W}) \,\, {\bar M}^{qp}_{\rho\gamma}
\left( {\bf S}_{\bf W} \right) \, S^{\gamma}_{W^{q}W^{p}} \,\,
d^{d} W $$
where
$${\bar M}^{qp}_{\rho\gamma} \left( {\bf S}_{\bf W} \right) 
\,\,\,\,\, \equiv \,\,\,\,\, \int_{0}^{1} \, M^{qp}_{\rho\gamma} 
\left( \lambda {\bf S}_{\bf W} \right) \,\, d \lambda $$

 To get a representation of the required form let us first note that
the 1-form (\ref{qOneForm}) is evidently invariant under the
transformations
\begin{equation}
\label{CrhoTrans}
S^{\rho} ({\bf X}) \quad \rightarrow \quad S^{\rho} ({\bf X})
\,\,\, + \,\,\, C^{\rho} \quad , \quad \quad 
C^{\rho} \,\, = \,\, {\rm const} 
\end{equation}

 As a corollary, we can claim that the values
$\, M^{qp}_{\alpha\gamma} ({\bf S}_{\bf X}) \, 
S^{\gamma}_{X^{q}X^{p}} \, $
can be also represented in the form
$$M^{qp}_{\alpha\gamma} ({\bf S}_{\bf X}) \, S^{\gamma}_{X^{q}X^{p}}
\,\,\,\,\, \equiv \,\,\,\,\, {\delta \over \delta S^{\alpha} ({\bf X})}
\, \int \, \left( S^{\rho} ({\bf W}) \, + \, C^{\rho} \right) \, 
{\bar M}^{qp}_{\rho\gamma} \left( {\bf S}_{\bf W} \right) \, 
S^{\gamma}_{W^{q}W^{p}} \,\, d^{d} W $$
for arbitrary values of $\, C^{\rho}$.

 We can claim then, that all the functionals
$$M_{(\rho)} \,\,\,\,\, = \,\,\,\,\, \int \,
{\bar M}^{qp}_{\rho\gamma} \left( {\bf S}_{\bf W} \right) \, 
S^{\gamma}_{W^{q}W^{p}} \,\, d^{d} W $$
have identically zero variation derivatives on the space of functions
$\, (S^{1} ({\bf X}), \dots, S^{m} ({\bf X}))$.

 According to the classical theorem (see e.g. \cite{Olver},
Chpt. 4, Thm. 4.7), any density 
$$\sigma_{(\rho)} ({\bf W}) \,\,\,\,\, = \,\,\,\,\, 
{\bar M}^{qp}_{\rho\gamma} \left( {\bf S}_{\bf W} \right) \,
S^{\gamma}_{W^{q}W^{p}} $$
can then be represented as the full divergence of the vector field
$\, {\bf v}_{(\rho)} ({\bf W}) \, $:
$${\bar M}^{qp}_{\rho\gamma} \left( {\bf S}_{\bf W} \right) \,
S^{\gamma}_{W^{q}W^{p}} \,\,\,\,\, \equiv \,\,\,\,\,
\left[ v^{r}_{(\rho)} ({\bf W}) \right]_{W^{r}} $$
where the values $\, v^{r}_{(\rho)} ({\bf W}) \, $ have in
general the form
$$v^{r}_{(\rho)} ({\bf W}) \,\,\,\,\,\,\,\, \equiv \,\,\,\,\,\,\,\,
S^{\mu} ({\bf W}) \,\, V^{rqp}_{(\rho)\mu\gamma} 
\left( {\bf S}_{\bf W} \right) \, S^{\gamma}_{W^{q}W^{p}} 
\,\,\,\,\, + \,\,\,\,\, 
u^{r}_{(\rho)} \left( {\bf S}_{\bf W} \right) $$
 
 The vector fields $\, \bm{\xi}_{(\rho\mu)} \, $ given by the
components
$$\xi^{r}_{(\rho\mu)} ({\bf W}) \,\,\,\,\, = \,\,\,\,\,
V^{rqp}_{(\rho)\mu\gamma} \left( {\bf S}_{\bf W} \right) \, 
S^{\gamma}_{W^{q}W^{p}} $$
represent divergence-free vector fields, so we can write according
to Lemma 2.1:
\begin{equation}
\label{VfRel}
V^{rqp}_{(\rho)\mu\gamma} \left( {\bf S}_{\bf W} \right) \,
S^{\gamma}_{W^{q}W^{p}} \,\,\,\,\,\,\,\, \equiv \,\,\,\,\,\,\,\,
\sum_{s \neq r} \, \left[ \, f^{sr}_{(\rho\mu)} 
\left( {\bf S}_{\bf W} \right) \right]_{W^{s}} 
\end{equation}
\begin{equation}
\label{SkewfRel}
f^{sr}_{(\rho\mu)} \left( {\bf S}_{\bf W} \right) 
\,\,\,\,\, \equiv \,\,\,\,\, - \, 
f^{rs}_{(\rho\mu)} \left( {\bf S}_{\bf W} \right)
\end{equation}
for some functions $\, f^{sr}_{(\rho\mu)} ({\bf S}_{\bf W})$.

 Using relations (\ref{VfRel}) - (\ref{SkewfRel}) we now easily get 
the relations
\begin{multline*}
\left[ \, S^{\mu} ({\bf W}) \,\, V^{rqp}_{(\rho)\mu\gamma}
\left( {\bf S}_{\bf W} \right) \, S^{\gamma}_{W^{q}W^{p}} 
\right]_{W^{r}} \,\,\,\,\, \equiv   \\
\equiv \,\,\,\,\, 
\left[ \, S^{\mu} ({\bf W}) \, \sum_{s \neq r} \, 
\left[ \, f^{sr}_{(\rho\mu)} \left( {\bf S}_{\bf W} \right) 
\right]_{W^{s}} \right]_{W^{r}} \,\,\,\,\, \equiv \,\,\,\,\,
S^{\mu}_{W^{r}} \, \sum_{s \neq r} \, \left[ \, 
f^{sr}_{(\rho\mu)} \left( {\bf S}_{\bf W} \right) \right]_{W^{s}} 
\end{multline*}
and
$$S^{\rho} ({\bf W}) \,\, {\bar M}^{qp}_{\rho\gamma}  
\left( {\bf S}_{\bf W} \right) \, S^{\gamma}_{W^{q}W^{p}} 
\,\,\,\,\,\,\,\, \equiv  \,\,\,\,\,\,\,\, 
S^{\rho} ({\bf W}) \, S^{\mu}_{W^{r}} \, \sum_{s \neq r} \, \left[ 
\, f^{sr}_{(\rho\mu)} \left( {\bf S}_{\bf W} \right) \right]_{W^{s}}
\,\,\,\,\, + \,\,\,\,\, S^{\rho} ({\bf W}) \, \left[ \,
u^{r}_{(\rho)} \left( {\bf S}_{\bf W} \right) \right]_{W^{r}} $$

 Finally, using integration by parts, we can claim that the
functional
$$H \,\,\,\,\, \equiv \,\,\,\,\, \int 
S^{\rho} ({\bf W}) \,\, {\bar M}^{qp}_{\rho\gamma}
\left( {\bf S}_{\bf W} \right) \, S^{\gamma}_{W^{q}W^{p}}
\,\, d^{d} W $$
can be represented in the form
$$H \,\,\,\,\, \equiv \,\,\,\,\, - \, \int \, \Big[ \,
S^{\rho}_{W^{s}} \, S^{\mu}_{W^{r}} \, \sum_{s \neq r} \, 
f^{sr}_{(\rho\mu)} \left( {\bf S}_{\bf W} \right) 
\,\,\,\,\, + \,\,\,\,\, S^{\rho}_{W^{r}} \,\,
u^{r}_{(\rho)} \left( {\bf S}_{\bf W} \right) \Big]
\,\, d^{d} W $$
which gives the proof of the lemma.

{\hfill Lemma 2.2 is proved.}

\vspace{0.3cm}

 Proof of Theorem 2.1.

 Using the homotopy operator approach for the closed 2-form 
(\ref{OprFormJ}) we obtain the relations
$$J_{\alpha\beta} [{\bf S}] \, ({\bf X}, {\bf Y}) 
\,\,\,\,\, = \,\,\,\,\,
{\delta \over \delta S^{\beta} ({\bf Y})} \,\, q_{\alpha} ({\bf X})
\,\,\, - \,\,\,
{\delta \over \delta S^{\alpha} ({\bf X})} \,\, q_{\beta} ({\bf Y}) $$
where
\begin{multline*}
q_{\alpha} ({\bf X}) \,\,\,\,\, \equiv \,\,\,\,\,
\int_{0}^{1} \, d \lambda \, \int \lambda \,
J_{\alpha\rho} [\lambda {\bf S}] \, ({\bf X}, {\bf W}) \,\,
S^{\rho} ({\bf W}) \,\, d^{d} W \,\,\,\,\, \equiv   \\
\equiv \,\,\,\,\, S^{\rho}_{X^{p}} \, \int_{0}^{1} \,
d \lambda \, \lambda \,\, \Omega^{p}_{\alpha\rho} \left(
\lambda {\bf S}_{\bf X} \right)  \,\,\,\,\, + \,\,\,\,\,
S^{\gamma}_{X^{p}X^{q}} \, S^{\rho} ({\bf X}) \,
\int_{0}^{1} \, d \lambda \, \lambda^{2} \,\,\,
\Gamma^{pq}_{\alpha\rho\gamma} \left( \lambda {\bf S}_{\bf X} \right)  
\end{multline*}

 We can see now that the coordinate change
\begin{equation}
\label{intermedtrans}
{\tilde Q}_{\alpha} ({\bf X}) \quad \rightarrow \quad
{\tilde Q}_{\alpha} ({\bf X}) \,\,\, + \,\,\, q_{\alpha} ({\bf X})
\end{equation}
gives the required form for the bracket (\ref{AdmBr}). However, we can 
see also that the transformation (\ref{intermedtrans}) does not have
the required form (\ref{ChangeCoordQ}). To get a transformation
of the form (\ref{ChangeCoordQ}) let us note again that the 2-form
(\ref{OprFormJ}) is evidently invariant under the transformations
(\ref{CrhoTrans}). As a consequence, we easily get then that any
transformation (\ref{CrhoTrans}), applied to the set
$\, \{ q_{\alpha} ({\bf X}) \} \, $,  gives a set of functions
$\, \{ q^{\prime}_{\alpha} ({\bf X}) \} \, $ having the property
that the change
$${\tilde Q}_{\alpha} ({\bf X}) \quad \rightarrow \quad
{\tilde Q}_{\alpha} ({\bf X}) \,\,\, + \,\,\, 
q^{\prime}_{\alpha} ({\bf X}) $$
transforms the bracket (\ref{AdmBr}) to the canonical form
(\ref{StandPBrCanForm}). It's not difficult to see, that this
circumstance means in fact that all the functions
$$\bm{\omega}_{(\rho)} ({\bf X}) \,\,\,\, = \,\,\,
\left(\omega_{(\rho) 1} ({\bf X}), \, \dots , \,
\omega_{(\rho) m} ({\bf X}) \right) $$
defined as
$$\omega_{(\rho) \alpha} ({\bf X}) \,\,\,\,\, \equiv \,\,\,\,\,
S^{\gamma}_{X^{p}X^{q}} \, \int_{0}^{1} \,
d \lambda \, \lambda^{2} \,\,\, \Gamma^{pq}_{\alpha\rho\gamma} 
\left( \lambda {\bf S}_{\bf X} \right) $$
represent coefficients of closed 1-forms on the space
$\, (S^{1} ({\bf X}), \dots, S^{m} ({\bf X})) \, $:
$${\delta \omega_{(\rho) \alpha} ({\bf X}) \over 
\delta S^{\beta} ({\bf Y})} \,\,\,\,\, \equiv \,\,\,\,\,
{\delta \omega_{(\rho) \beta} ({\bf Y}) \over
\delta S^{\alpha} ({\bf X})} \quad , \quad \quad
\rho \, = \, 1, \dots, m $$

 Using Lemma 2.2 we can claim then that the functions
$\, \bm{\omega}_{(\rho)} ({\bf X}) \, $ can be locally represented
in the form
$$\omega_{(\rho) \alpha} ({\bf X}) \,\,\,\,\, \equiv \,\,\,\,\,
{\delta \over \delta S^{\alpha} ({\bf X})} \, \int \,
h_{(\rho)} \left( {\bf S}_{\bf W} \right) \,\, d^{d} W $$
for some functions $\, h_{(\rho)} ({\bf S}_{\bf W})$.

 Let us put now
$${\tilde q}_{\alpha} ({\bf X}) \,\,\,\,\,\, = \,\,\,\,\,\,
q_{\alpha} ({\bf X}) \,\,\,\,\, - \,\,\,\,\, 
{\delta \over \delta S^{\alpha} ({\bf X})} \, \int \,
S^{\rho} ({\bf W}) \,\,
h_{(\rho)} \left( {\bf S}_{\bf W} \right) \,\, d^{d} W $$
and define
\begin{equation}
\label{QalphaCoord}
Q_{\alpha} ({\bf X}) \,\,\,\,\, = \,\,\,\,\,
{\tilde Q}_{\alpha} ({\bf X}) \,\,\, + \,\,\,
{\tilde q}_{\alpha} ({\bf X})
\end{equation}

 It can be seen now that the coordinate change (\ref{QalphaCoord})
has the necessary form (\ref{ChangeCoordQ}). Besides that, the 
change (\ref{QalphaCoord}) transforms the bracket (\ref{AdmBr})
to the canonical form like the transformation (\ref{intermedtrans}).

{\hfill Theorem 2.1 is proved.}

\vspace{0.3cm}

 As a corollary of Theorems 2.1 - 2.1$^{\prime}$ we can claim that 
any system (\ref{MultDimConsWhithSyst}) with $\, s \, = \, 0 \, $
can be written locally in the Lagrangian form
$$\delta \, \int \, \Big[ \, Q_{\alpha} ({\bf X}) \, S^{\alpha}_{T}
\,\,\, - \,\,\, \langle P_{H} \rangle \left( {\bf S}_{X^{1}}, \dots ,
{\bf S}_{X^{d}}, \, {\bf Q} ({\bf X}) \right) \Big] \,\,
d^{d} X \, d T  \quad  =  \quad  0  $$
after the transition to the variables
$\, ( S^{1} ({\bf X}) , \dots , S^{m} ({\bf X}) , \,
Q_{1} ({\bf X}) , \dots , Q_{m} ({\bf X}) ) \, $.

 In the non-degenerate case, when the values 
$\, Q_{\alpha} ({\bf X}) \, $ can be expressed in terms of   \linebreak
$\, ( {\bf S}_{T}, {\bf S}_{X^{1}}, \dots , {\bf S}_{X^{d}} ) \, $
from the first part of system (\ref{MultDimConsWhithSyst}), we can
write system (\ref{MultDimConsWhithSyst}) in the   \linebreak
``standard'' Lagrangian form
$$\delta \, \int \, \left[ \, \sum_{\alpha=1}^{m} \left(
S^{\alpha}_{T} \right)^{2} \,\,\, - \,\,\, \langle P_{H} \rangle 
\left( {\bf S}_{T}, {\bf S}_{X^{1}}, \dots , {\bf S}_{X^{d}} \right)
\right] \, d^{d} X \, d T  \quad  =  \quad  0  $$

 It is not difficult to see, that we have to require the
non-degeneracy conditions
$${\rm det} \,\, \left| \left| \,\,\, \left.
{\partial \omega^{\alpha} \over \partial Q_{\beta}} 
\right|_{{\bf S}_{\bf X}} \, \right| \right|  \quad  =  \quad
{\rm det} \,\, \left| \left| \,\,\, \left. {\partial^{2} 
\langle P_{H} \rangle \over \partial Q_{\alpha} \partial Q_{\beta}}
\right|_{{\bf S}_{\bf X}} \, \right| \right|  \quad  \neq  \quad 0 $$
in this situation.

\vspace{0.3cm}

 At the end of the chapter let us discuss the group of canonical
transformations for the bracket (\ref{StandPBrCanForm}) having the
``physical'' form 
\begin{equation}
\label{QQprimeCanTrans}
Q^{\prime}_{\alpha} ({\bf X}) \,\,\,\,\, = \,\,\,\,\,
Q_{\alpha} ({\bf X}) \,\,\, + \,\,\,
q_{\alpha} \left( {\bf S}_{X^{1}}, \dots, {\bf S}_{X^{d}} \right)
\end{equation}

 It is easy to see that the transformation (\ref{QQprimeCanTrans})
represents a canonical transformation for the bracket
(\ref{StandPBrCanForm}) if and only if we have the identities
$${\delta \over \delta S^{\beta} ({\bf Y})} \,\,\,
q_{\alpha} \left( {\bf S}_{\bf X} \right) \,\,\,\,\, - \,\,\,\,\,
{\delta \over \delta S^{\alpha} ({\bf X})} \,\,\, 
q_{\beta} \left( {\bf S}_{\bf Y} \right) \quad  \equiv  \quad  0 $$

 It is easy to check also that the identities above are equivalent
to the following relations
\begin{equation}
\label{qalphaqbetaRel1}
{\partial q_{\alpha} \over \partial S^{\beta}_{X^{q}}}
\,\,\, \equiv \,\,\, - \,
{\partial q_{\beta} \over \partial S^{\alpha}_{X^{q}}}
\quad , \quad \quad \quad
{\partial^{2} q_{\beta} \over \partial S^{\alpha}_{X^{q}}
\partial S^{\gamma}_{X^{p}}} \,\,\, + \,\,\,
{\partial^{2} q_{\beta} \over \partial S^{\alpha}_{X^{p}}
\partial S^{\gamma}_{X^{q}}} \,\,\, \equiv \,\,\, 0 
\end{equation}

 From (\ref{qalphaqbetaRel1}) we easily get also the following 
relations
$${\partial q_{\alpha} \over \partial S^{\alpha}_{X^{q}}}
\,\,\, \equiv \,\,\, 0  \quad , \quad \quad \quad
{\partial^{2} q_{\beta} \over \partial S^{\alpha}_{X^{q}}
\partial S^{\alpha}_{X^{p}}} \,\,\, \equiv \,\,\, 0  
\quad , \quad \quad \quad
{\partial^{2} q_{\beta} \over \partial S^{\alpha}_{X^{q}}
\partial S^{\gamma}_{X^{q}}} \,\,\, \equiv \,\,\, 0 $$
(no summation).

 It's not difficult to see that the group of the canonical
transformations (\ref{QQprimeCanTrans}) represents in fact
a finite-dimensional linear space with the basis elements, which
can be described in the following way:

Consider all possible pairs of sets 
$\, ({\cal M}_{1}, \, {\cal M}_{2} ) \, $:
$${\cal M}_{1} \,\,\, = \,\,\, \left( \alpha_{1}, \, \dots , \,
\alpha_{l+1} \right)  \quad , \quad \quad 
\alpha_{j} \, \in \, \{ 1, \, \dots , \, m \} \quad , \quad \quad
\alpha_{1} \, < \, \alpha_{2} \, < \, \dots \, < \, \alpha_{l+1} $$
$${\cal M}_{2} \,\,\, = \,\,\, \left( q_{1}, \, \dots , \, 
q_{l} \right)  \quad , \quad \quad
q_{j} \, \in \, \{ 1, \, \dots , \, d \} \quad , \quad \quad
q_{1} \, < \, q_{2} \, < \, \dots \, < \, q_{l} $$
for all possible $\, l \, \geq \, 0 \, $.

 Consider the functions 
$\, {\bf q}_{\, ({\cal M}_{1}, \, {\cal M}_{2})} \, $ having the form
$$q_{\alpha \, ({\cal M}_{1}, \, {\cal M}_{2})}  \quad  =  \quad
\begin{cases}
\quad \quad  0  \quad \quad \,\,\,\, , \quad \quad 
\alpha \, \notin \, {\cal M}_{1}  \\
\Delta^{\alpha}_{({\cal M}_{1}, \, {\cal M}_{2})} 
\quad , \quad \quad \alpha \, \in \, {\cal M}_{1}
\end{cases}  $$
where
$$\Delta^{\alpha}_{({\cal M}_{1}, \, {\cal M}_{2})}
\quad  \equiv  \quad  (-1)^{j-1} \,\,\, {\rm det} \,\,
\begin{Vmatrix}
S^{\alpha_{1}}_{X^{q_{1}}}  &  S^{\alpha_{1}}_{X^{q_{2}}}  &
\dots  &  S^{\alpha_{1}}_{X^{q_{l}}}  \\
\dots  &  \dots  &  \dots  &  \dots   \\
{\hat S}^{\alpha_{j}}_{X^{q_{1}}}  &  
{\hat S}^{\alpha_{j}}_{X^{q_{2}}}  &
\dots  &  {\hat S}^{\alpha_{j}}_{X^{q_{l}}}  \\
\dots  &  \dots  &  \dots  &  \dots   \\
S^{\alpha_{l+1}}_{X^{q_{1}}}  &  S^{\alpha_{l+1}}_{X^{q_{2}}}  &
\dots  &  S^{\alpha_{l+1}}_{X^{q_{l}}}  
\end{Vmatrix}
\,\,\,  ,  \quad \,  \alpha \, = \, \alpha_{j} \, \in \, 
{\cal M}_{1}  \,\,\,  ,  \quad \,  l \, \geq \, 1  $$
(the hats mean that the corresponding row is absent in the
matrix).

 Let us also put by definition
$$\Delta^{\alpha}_{({\cal M}_{1}, \, {\cal M}_{2})}
\quad  \equiv  \quad  1 $$
for $\, l \, = \, 0 \, $, 
$\,\, {\cal M}_{1} \, = \, \{ \alpha \} \, $, 
$\,\, {\cal M}_{2} \, = \, \varnothing \, $.

 The functions 
$\, \{ {\bf q}_{\, ({\cal M}_{1}, \, {\cal M}_{2})} \} \, $  
can be considered now as the basis of the linear space, representing
the group of the canonical transformations (\ref{QQprimeCanTrans}).

\section{On more complicated (pseudo-)Canonical forms.}
\setcounter{equation}{0}

 In this chapter we will consider brackets (\ref{AveragedBracket})
in the case of presence of additional parameters
$\, (n^{1}, \dots , n^{s}) , \, $ connected with the presence of
annihilators of the initial bracket (\ref{MultDimPBr}). As we said
in Introduction, we will assume here that the bracket
(\ref{AveragedBracket}) is obtained by the averaging of the bracket
(\ref{MultDimPBr}) on a complete Hamiltonian family 
$\, \Lambda \, $ of $m$-phase solutions of system (\ref{EvInSyst}),
equipped with a minimal set of commuting integrals, which implies,
in particular, that the number of the parameters
$\, (n^{1}, \dots , n^{s}) \, $ is exactly equal to the number of
annihilators of the bracket (\ref{MultDimPBr}) on the space of
quasiperiodic functions.

 As we will see below, the canonical form of the bracket
(\ref{AveragedBracket}) should be understood in this case in more
general sense and represents in fact the separation of the
``standard'' canonical variables
$$\left( S^{1} ({\bf X}), \dots , S^{m} ({\bf X}), \,
Q_{1} ({\bf X}), \dots , Q_{m} ({\bf X}) \right) $$
and some special variables
$$\left( {\bar N}^{1} ({\bf X}), \dots , {\bar N}^{s} ({\bf X}) 
\right) $$
with their own Poisson bracket.

 As in the previous chapter, we consider here
the transformations of the ``physical'' variables
$\, (U^{1} ({\bf X}), \dots , U^{m+s} ({\bf X})) \, $
having the form
$$U^{\prime \gamma} ({\bf X})  \quad = \quad  U^{\prime \gamma}
\left( {\bf S}_{X^{1}}, \dots , {\bf S}_{X^{d}}, \,
{\bf U}({\bf X}) \right) \,\,\, , \quad \quad
\gamma \, = \, 1, \dots, m + s \,\,\, , $$
which can be called the transformations of
``Hydrodynamic Type''. As above, the variables  \linebreak
$\, ( S^{1} ({\bf X}), \dots , S^{m} ({\bf X}) ) \, $
will be always considered here as the first part of canonical 
variables for every bracket (\ref{AveragedBracket}).

 Let us consider now a special class of the Poisson brackets
(\ref{MultDimPBr}) having some special ``physical'' property.

\vspace{0.3cm}

\noindent
{\bf Definition 3.1.}

{\it Let us say that the bracket (\ref{MultDimPBr}) has
annihilators of the physical form if all the independent
annihilators of (\ref{MultDimPBr}) on the space of quasiperiodic 
functions can be represented in the form: 
$$C^{l}  \,\,\, = \,\,\,  \int \, c^{l} \left( \bm{\varphi}, \,
\bm{\varphi}_{\bf x}, \, \dots \right) \, d^{d} x \,\,\,\,\, ,
\quad \quad \quad  l \, = \, 1, \dots , s $$
with some smooth functions
$\, c^{l} \, ( \bm{\varphi}, \, \bm{\varphi}_{\bf x}, \, \dots )$. 
}

\vspace{0.3cm}

 In particular, for a complete Hamiltonian family $\, \Lambda \, $
of $m$-phase solutions of system (\ref{EvInSyst}) the Definition 3.1 
requires that the functions
$${\bf v}^{(l)}_{[{\bf a}, \bm{\theta}_{0}]} ({\bf x}) \,\,\, = \,\,\,
\left( v^{(l)}_{[{\bf a}, \bm{\theta}_{0}] 1} ({\bf x}) , \, \dots ,
\, v^{(l)}_{[{\bf a}, \bm{\theta}_{0}] n} ({\bf x}) \right) \,\,\, , \,\,   
\quad  \quad
v^{(l)}_{[{\bf a}, \bm{\theta}_{0}] i} ({\bf x}) \,\,\, = \,\,\,
\left. {\delta C^{(l)} \over \delta \varphi^{i} ({\bf x})}
\right|_{\Lambda_{{\bf k}_{1}, \dots , {\bf k}_{d}}} \,\,\, ,   
\quad \,\,\,  l \, = \, 1, \dots , s  $$  
represent the full basis of solutions of system (\ref{AnnSolCoordRepr}),
having the form (\ref{AnnVarDerForm}), everywhere on $\, \Lambda \, $
in accordance with the Definition 1.2.

\vspace{0.3cm}

 Let us come back now to the variables
$$\left( S^{1} ({\bf X}), \dots , S^{m} ({\bf X}), \,
{\tilde Q}_{1} ({\bf X}), \dots ,
{\tilde Q}_{m} ({\bf X}), \, {\tilde N}^{1} ({\bf X}), \dots ,
{\tilde N}^{s} ({\bf X}) \right) $$
for the bracket (\ref{AveragedBracket}), introduced in the
previous chapter. We can formulate here the following lemma:

\vspace{0.3cm}

\noindent
{\bf Lemma 3.1.}

{\it Let the bracket (\ref{MultDimPBr}) have annihilators of the
physical form and the family $\, \Lambda \, $ represent a complete
Hamiltonian family of $m$-phase solutions of system (\ref{EvInSyst})
equipped with a minimal set of commuting integrals. Let the bracket
(\ref{AveragedBracket}) represent the averaging of the bracket
(\ref{MultDimPBr}) on the family $\, \Lambda $.

 Then the variables
$$\left( {\tilde Q}_{1} ({\bf X}), \dots ,
{\tilde Q}_{m} ({\bf X}), \, {\tilde N}^{1} ({\bf X}), \dots ,
{\tilde N}^{s} ({\bf X}) \right) $$
can be chosen in the form
$$\left( {\tilde Q}_{1} ({\bf X}), \dots ,
{\tilde Q}_{m} ({\bf X}), \, N^{1} ({\bf X}), \dots ,
N^{s} ({\bf X}) \right) $$
where
$$N^{l} \quad \equiv \quad \langle c^{l} \rangle \quad \equiv \quad
\int_{0}^{2\pi}\!\!\!\dots\int_{0}^{2\pi} \, c^{l} \left( \bm{\Phi},
\, k^{\alpha_{1}}_{1} \bm{\Phi}_{\theta^{\alpha_{1}}}, \, \dots ,
\, k^{\alpha_{d}}_{d} \bm{\Phi}_{\theta^{\alpha_{d}}} , \, \dots 
\right) \,
{d^{m} \theta \over (2\pi)^{m}} $$
represent the averaged densities of annihilators.
}

\vspace{0.3cm}

 Proof.

 Let the set $\, \{ I^{1}, \dots , I^{m+s} \} \, $ represent a
minimal set of commuting integrals (\ref{IgammaForm}) for the
family $\, \Lambda $. Let us assume without lost of generality 
that we have
$${\rm rk} \,\, \left| \left| \, \omega^{\alpha \gamma} \,
({\bf k}_{1}, \dots, {\bf k}_{d}, \, {\bf U}) \, \right| \right|
\,\,\, = \,\,\, m \,\,\, , \quad \quad
\gamma \, = \, 1, \dots , m $$
for the corresponding frequencies $\, \bm{\omega}^{\gamma} $.

 It's not difficult to see that the variation derivatives
$$\left( {\delta I^{1} \over \delta \varphi^{i} ({\bf x}) } \, , \,
\dots , \, {\delta I^{m} \over \delta \varphi^{i} ({\bf x}) } 
\, , \,\, {\delta C^{1} \over \delta \varphi^{i} ({\bf x}) } 
\, , \, \dots , \, 
{\delta C^{s} \over \delta \varphi^{i} ({\bf x}) } \right) $$
represent in this case a linearly independent system
on $\, \Lambda , \, $ so the 
values
$$U^{\gamma} \,\,\, = \,\,\, \langle P^{\gamma} \rangle \,\,\, ,
\,\,\,\,\, \gamma \, = \, 1, \dots , m \,\,\,  , \quad \quad
N^{l} \,\,\, = \,\,\, \langle c^{l} \rangle \,\,\, ,
\,\,\,\,\, l \, = \, 1, \dots , s \,\,\, , $$
give a set of independent parameters on every family
$\, \Lambda_{{\bf k}_{1}, \dots, {\bf k}_{d}}$. We can easily see 
then that the functionals
$$\left( I^{1}, \, \dots , \, I^{m} , \,\, C^{1}, \, \dots , \,
C^{s} \right) $$
represent a minimal set of commuting integrals for the family
$\, \Lambda $, satisfying all the requirements of Definition 1.3.

 Consider now the bracket (\ref{AveragedBracket}) in the coordinates
$$\left( S^{1} ({\bf X}), \dots , S^{m} ({\bf X}), \,
U^{1} ({\bf X}), \dots , U^{m} ({\bf X}), \, 
N^{1} ({\bf X}), \dots , N^{s} ({\bf X}) \right) $$

  Using the Jacobi identities
$$\left\{ \left\{ U^{\gamma} ({\bf X}) \, , \, S^{\alpha} ({\bf Y}) 
\right\} \, , \, S^{\beta} ({\bf Z}) \right\} \,\, - \,\,
\left\{ \left\{ U^{\gamma} ({\bf X}) \, , \, S^{\beta} ({\bf Z})
\right\} \, , \, S^{\alpha} ({\bf Y}) \right\}
\,\,\, \equiv \,\,\, 0 \,\,\, , \quad \quad 
\gamma \, = \, 1, \dots , m \,\,\, , $$
we easily get that the vector fields
$${\vec \xi}_{(\alpha)} \,\, = \,\,
\left( \, \omega^{\alpha \, 1}
({\bf k}_{1}, \dots, {\bf k}_{d}, \,  {\bf U}, \, {\bf N} ) , \,
\dots, \, \omega^{\alpha \, m}
({\bf k}_{1}, \dots, {\bf k}_{d}, \,  {\bf U}, \, {\bf N} ) 
\, \right)^{t} $$
represent commuting vector fields, tangent to the submanifolds
$\, ({\bf k}_{1}, \dots, {\bf k}_{d}, \, {\bf N}) 
\, = \, {\rm const}$.

 Since the set $\, \{ {\vec \xi}_{(\alpha)} \} \, $ is linearly
independent we can claim again that we can choose the variables
$$\left( {\tilde Q}_{1} 
({\bf k}_{1}, \dots, {\bf k}_{d}, \,  {\bf U}, \, {\bf N} ) , \,
\dots, \, {\tilde Q}_{m}
({\bf k}_{1}, \dots, {\bf k}_{d}, \,  {\bf U}, \, {\bf N} ) 
\right) $$
on each submanifold, such that the vectors 
$\, {\vec \xi}_{(\alpha)} \, $ get the coordinate representation:
$${\vec \xi}_{(\alpha)} \, = \, 
(0, \, \dots , \, 1 , \, \dots , \, 0)$$

 Easy to see that we get now the required coordinate system using
the variables   \linebreak
$\, (N^{1} ({\bf X}), \, \dots , \, N^{s} ({\bf X})) \, $ and
$${\tilde Q}_{\alpha} ({\bf X}) \,\,\, = \,\,\,
{\tilde Q}_{\alpha} \left( {\bf S}_{X^{1}}, \dots , {\bf S}_{X^{d}},
\, {\bf U} ({\bf X}) , \, {\bf N} ({\bf X}) \right) 
\,\,\, , \quad \quad   \alpha \, = \, 1, \dots , m $$

{\hfill Lemma 3.1 is proved.}

\vspace{0.3cm}

\noindent
{\bf Lemma 3.2.}

{\it Let the bracket (\ref{MultDimPBr}) have annihilators of the
physical form and the family $\, \Lambda \, $ represent a complete
Hamiltonian family of $m$-phase solutions of system (\ref{EvInSyst})
equipped with a minimal set of commuting integrals. Let the bracket
(\ref{AveragedBracket}) represent the averaging of the bracket
(\ref{MultDimPBr}) on the family $\, \Lambda $.

 Consider the variables on the space $\, {\bf U} ({\bf X}) \, $
introduced in Lemma 3.1.

 Then the bracket (\ref{AveragedBracket}) has in the variables
$$\left( S^{1} ({\bf X}), \dots , S^{m} ({\bf X}), \,
{\tilde Q}_{1} ({\bf X}), \dots ,
{\tilde Q}_{m} ({\bf X}), \, N^{1} ({\bf X}), \dots ,
N^{s} ({\bf X}) \right) $$
the form

$$\left\{ S^{\alpha} ({\bf X}) \, , \, S^{\beta} ({\bf Y}) \right\}
\,\,\, = \,\,\, 0  $$
$$\left\{ S^{\alpha} ({\bf X}) \, , \,
{\tilde Q}_{\beta} ({\bf Y}) \right\}
\,\,\, = \,\,\, \delta^{\alpha}_{\beta} \,\,\,
\delta ({\bf X} - {\bf Y}) \,\,\, , \,\,\,\,\,\,\,\,\,\,
\left\{ S^{\alpha} ({\bf X}) \, , \,
N^{l}  ({\bf Y}) \right\} \,\,\, = \,\,\, 0 $$
\begin{multline*}
\left\{ {\tilde Q}_{\alpha} ({\bf X}) \, , \,
{\tilde Q}_{\beta} ({\bf Y}) \right\} \,\,\, = 
\,\,\, \Omega_{\alpha\beta}^{p}
\left( {\bf S}_{\bf X}, {\bf N} \right) \,\,
\delta_{X^{p}} ({\bf X} - {\bf Y}) \,\,\, +   \\
+ \,\,\, \Gamma_{\alpha\beta\gamma}^{pq}
\left( {\bf S}_{\bf X}, {\bf N} \right) \,
S^{\gamma}_{X^{p}X^{q}} \,\, \delta ({\bf X} - {\bf Y}) \,\,\, + \,\,\,
\Pi_{\alpha\beta r}^{p} \left( {\bf S}_{\bf X}, {\bf N} \right) \,
N^{r}_{X^{p}} \,\, \delta ({\bf X} - {\bf Y}) 
\end{multline*}
$(\Gamma_{\alpha\beta\gamma}^{pq} \, \equiv \,
\Gamma_{\alpha\beta\gamma}^{qp})$,
\begin{equation}
\label{IntermedSQNBr}
\left\{ {\tilde Q}_{\alpha} ({\bf X}) \, , \,
N^{l}  ({\bf Y}) \right\}
\,\,\, = \,\,\, A_{\alpha}^{l,p}
\left( {\bf S}_{\bf X}, {\bf N} \right) \,\,
\delta_{X^{p}} ({\bf X} - {\bf Y}) 
\end{equation}
$$\left\{  N^{l}  ({\bf X}) \, , \, N^{k}  ({\bf Y}) \right\}
\,\,\, = \,\,\, g^{lk,p}
\left( {\bf S}_{\bf X}, {\bf N} \right) \,\,
\delta_{X^{p}} ({\bf X} - {\bf Y}) $$

}

\vspace{0.3cm}

 Proof.

 What we have actually to prove is the absence of the
``$\delta$ - terms'' in the last two expressions of 
(\ref{IntermedSQNBr}). To prove this fact let us first note that,
according to the definition of annihilators, the corresponding terms
are absent in the Poisson brackets of the densities
$\, P^{\gamma} (\bm{\varphi}, \, \bm{\varphi}_{\bf x}, \, \dots ) \, $  
and $\, C^{l} (\bm{\varphi}, \, \bm{\varphi}_{\bf x}, \, \dots ) \, $
with $\, C^{k} (\bm{\varphi}, \, \bm{\varphi}_{\bf y}, \, \dots ) \, $:  
$$\{ P^{\gamma} ({\bf x}) \, , \, C^{k} ({\bf y}) \} \,\,\, = \,\,\,
\sum_{l_{1},\dots,l_{d}} \, G^{\gamma k}_{l_{1} \dots l_{d}}
(\bm{\varphi}, \bm{\varphi}_{\bf x}, \dots ) \,\,\,   
\delta^{(l_{1})} (x^{1} - y^{1}) \, \dots \,
\delta^{(l_{d})} (x^{d} - y^{d}) $$
$$\{ C^{l} ({\bf x}) \, , \, C^{k} ({\bf y}) \} \,\,\, = \,\,\,
\sum_{l_{1},\dots,l_{d}} \, W^{l k}_{l_{1} \dots l_{d}}
(\bm{\varphi}, \bm{\varphi}_{\bf x}, \dots ) \,\,\,
\delta^{(l_{1})} (x^{1} - y^{1}) \, \dots \,
\delta^{(l_{d})} (x^{d} - y^{d}) $$
($l_{1}, \dots, l_{d} \geq 0 \, , \,\,\,$
$(l_{1}, \dots, l_{d}) \, \neq \, (0, \dots , 0)$).

 According to the averaging procedure we can claim then the absence
of the terms, containing $\, \delta \, ({\bf X} - {\bf Y}) \, , \, $ 
in the Poisson brackets 
$\, \{ U^{\gamma} ({\bf X}) \, , \, N^{k} ({\bf Y}) \} \, $ and
$\, \{ N^{l} ({\bf X}) \, , \, N^{k} ({\bf Y}) \} \, $ for the
bracket (\ref{AveragedBracket}). Easy to see then, that the same 
property is valid also for the brackets
$\, \{ {\tilde Q}_{\alpha} ({\bf X}) \, , \, N^{k} ({\bf Y}) \} \, $ 
and $\, \{ N^{l} ({\bf X}) \, , \, N^{k} ({\bf Y}) \} \, $ after the
transition to the variables
$ \, ( {\tilde Q}_{1} ({\bf X}), \dots ,
{\tilde Q}_{m} ({\bf X}), \, N^{1} ({\bf X}), \dots ,
N^{s} ({\bf X}) ) \, $.

{\hfill Lemma 3.2 is proved.}

\vspace{0.3cm}

 Let us note also here that from the form of the Poisson brackets
for the functionals $\, N^{k} ({\bf Y}) \, $ we can also conclude
that all the functionals
$$n^{l} \,\,\, = \,\,\, \int \, N^{l} ({\bf X}) \,\, d^{d} X $$
represent annihilators of the physical form for the averaged bracket
(\ref{AveragedBracket}).

\vspace{0.3cm}

 It's not difficult to check that just from the skew-symmetry of the
bracket (\ref{IntermedSQNBr}) we get the relations
$$g^{lk, p} \,\,\, \equiv \,\,\, g^{kl, p} \quad , \quad \quad  
\quad \left[ g^{kl, p} \right]_{X^{p}} \,\,\, \equiv \,\,\, 0 $$
 
 From the second relation above we then easily get the relations  
$${\partial g^{kl, p} \over \partial N^{r}} \,\,\, \equiv \,\,\, 0
\quad , \quad \quad \quad
{\partial g^{kl, p} \over \partial S^{\alpha}_{X^{q}}} \,\, + \,\,
{\partial g^{kl, q} \over \partial S^{\alpha}_{X^{p}}}
 \,\,\, \equiv \,\,\, 0 $$
for the functions
$\, g^{kl, p} \,
({\bf S}_{X^{1}}, \dots , {\bf S}_{X^{d}}, \, {\bf N})$.
 
 So, we can put now
$\, g^{kl, p} \,\, = \,\,  g^{kl, p} \,
({\bf S}_{X^{1}}, \dots , {\bf S}_{X^{d}}) \, $
for all the functions $\,  g^{kl, p} $. It can be also seen, that 
the second part of the relations above implies the relations
$${\partial^{2} g^{kl, p} \over \partial S^{\alpha}_{X^{q}}
\partial S^{\gamma}_{X^{r}}} \,\,\, \equiv \,\,\, - \,
{\partial^{2} g^{kl, p} \over \partial S^{\alpha}_{X^{r}} 
\partial S^{\gamma}_{X^{q}}} $$

 Like in Chapter 2, we can actually claim here that all the
functions
$${\bf g}^{kl} \,\,\, = \,\,\, {\bf g}^{lk} \,\,\, = \,\,\, 
\left( g^{kl, 1} , \, \dots , \, g^{kl, d} \right) $$
belong to a finite-dimensional linear space with the basis elements,
which can be described in the following way:

 Consider all possible pairs of sets
$\, ({\cal N}_{1}, \, {\cal N}_{2} ) \, $:
$${\cal N}_{1} \,\,\, = \,\,\, \left( \alpha_{1}, \, \dots , \,
\alpha_{l} \right)  \quad , \quad \quad
\alpha_{j} \, \in \, \{ 1, \, \dots , \, m \} \quad , \quad \quad
\alpha_{1} \, < \, \alpha_{2} \, < \, \dots \, < \, \alpha_{l} $$
$${\cal N}_{2} \,\,\, = \,\,\, \left( q_{1}, \, \dots , \,
q_{l+1} \right)  \quad , \quad \quad
q_{j} \, \in \, \{ 1, \, \dots , \, d \} \quad , \quad \quad
q_{1} \, < \, q_{2} \, < \, \dots \, < \, q_{l+1} $$
for all possible $\, l \, \geq \, 0 \, $.

 Consider the functions
$\, {\bf g}_{\, ({\cal N}_{1}, \, {\cal N}_{2})} \, $ having the form
$$g^{p}_{\, ({\cal N}_{1}, \, {\cal N}_{2})}  \quad  =  \quad
\begin{cases}
\quad \quad  0  \quad \quad \,\,\,\, , \quad \quad
p \, \notin \, {\cal N}_{2}  \\
{\bar \Delta}^{p}_{({\cal N}_{1}, \, {\cal N}_{2})}
\,\,\, \quad , \quad \quad p \, \in \, {\cal N}_{2}
\end{cases}  $$
where
$${\bar \Delta}^{p}_{({\cal N}_{1}, \, {\cal N}_{2})}
\quad  \equiv  \quad  (-1)^{j-1} \,\,\, {\rm det} \,\,
\begin{Vmatrix}
S^{\alpha_{1}}_{X^{p_{1}}}  &  \dots  &
{\hat S}^{\alpha_{1}}_{X^{p_{j}}}  & \dots  &
S^{\alpha_{1}}_{X^{p_{l+1}}}  \\
S^{\alpha_{2}}_{X^{p_{1}}}  & \dots  &
{\hat S}^{\alpha_{2}}_{X^{p_{j}}}  & \dots  &  
S^{\alpha_{2}}_{X^{p_{l+1}}}  \\
\vdots  &  \vdots  &  \vdots  &  \vdots  &  \vdots  \\
S^{\alpha_{l}}_{X^{p_{1}}}  &  \dots  &
{\hat S}^{\alpha_{l}}_{X^{p_{j}}}  & \dots  &  
S^{\alpha_{l}}_{X^{p_{l+1}}}
\end{Vmatrix}
\quad  ,  \quad \,\,\,\,\,  p \, = \, p_{j} \, \in \,
{\cal N}_{2}  \quad  ,  \quad \,\,\,\,\,
l \, \geq \, 1  $$
(the hats mean that the corresponding column is absent in the
matrix).

 We also put by definition
$${\bar \Delta}^{p}_{({\cal N}_{1}, \, {\cal N}_{2})}
\quad  \equiv  \quad  1 $$
for $\, l \, = \, 0 \, $,
$\,\, {\cal N}_{1} \, = \, \varnothing \, $,
$\,\, {\cal N}_{2} \, = \, \{ p \} \, $.
 
 So, we can write here
$${\bf g}^{kl} \,\,\,\,\, \in \,\,\,\,\, {\rm Span} \,
\left\{ {\bf g}_{\, ({\cal N}_{1}, \, {\cal N}_{2})} \right\} $$
for all $\,\, k, l \, = \, 1, \dots, d \, $.

\vspace{0.3cm}

\noindent
{\bf Definition 3.2.}

{\it Let the bracket (\ref{MultDimPBr}) have annihilators of the
physical form and the family $\, \Lambda \, $ represent a complete  
Hamiltonian family of $m$-phase solutions of system (\ref{EvInSyst})
equipped with a minimal set of commuting integrals. Let the bracket
(\ref{AveragedBracket}) represent the averaging of the bracket
(\ref{MultDimPBr}) on the family $\, \Lambda $.

1) We say that the bracket (\ref{AveragedBracket}) has a 
non-degenerate annihilator part if we have 
$${\rm det} \, \left| \left| g^{lk, p} \right| \right| 
\,\,\, \neq \,\,\, 0 $$
at least for one $\, p \, $ in coordinates 
$\, ( S^{1} ({\bf X}), \dots , S^{m} ({\bf X}), \,
{\tilde Q}_{1} ({\bf X}), \dots , 
{\tilde Q}_{m} ({\bf X}), \, N^{1} ({\bf X}), \dots ,
N^{s} ({\bf X}) ) $.

2) We say that the bracket (\ref{AveragedBracket}) has a
simple annihilator part if we have
$$g^{lk, p} \,\,\, = \,\,\, {\rm const} $$ 
(all $p$) in coordinates
$\, ( S^{1} ({\bf X}), \dots , S^{m} ({\bf X}), \,
{\tilde Q}_{1} ({\bf X}), \dots ,
{\tilde Q}_{m} ({\bf X}), \, N^{1} ({\bf X}), \dots ,
N^{s} ({\bf X}) ) $.

}

\vspace{0.3cm}

 Let us formulate now the Theorem related to the canonical form
of the brackets (\ref{AveragedBracket}), which are obtained by the
averaging of the brackets (\ref{MultDimPBr}) having the special
property, formulated above.

\vspace{0.3cm}

\noindent
{\bf Theorem 3.1.}

{\it Let the bracket (\ref{MultDimPBr}) have annihilators of the
physical form and the family $\, \Lambda \, $ represent a complete
Hamiltonian family of $m$-phase solutions of system (\ref{EvInSyst})
equipped with a minimal set of commuting integrals. Let the bracket
(\ref{AveragedBracket}) represent the averaging of the bracket
(\ref{MultDimPBr}) on the family $\, \Lambda $.

 Let the bracket (\ref{AveragedBracket}) have a simple 
non-degenerate annihilator part.

 Then there exists locally a smooth change of coordinates
\begin{equation}
\label{UQNTrans}
\begin{array}{c}
\left( U^{1}, \dots, U^{m+s} \right) \,\,\, \rightarrow \,\,\,
\left( Q_{1}, \dots, Q_{m}, \, 
{\bar N}^{1}, \dots, {\bar N}^{s} \right)    \\  \\
Q_{\alpha} \,\, = \,\, Q_{\alpha} \left(
{\bf S}_{X^{1}}, \dots, {\bf S}_{X^{d}}, \,
{\bf U} \right) \,\,\,\,\, , \,\,\,\,\,\,\,\,
{\bar N}^{l} \,\, = \,\,  {\bar N}^{l} \left(
{\bf S}_{X^{1}}, \dots, {\bf S}_{X^{d}}, \, {\bf U} \right) 
\,\,\, , 
\end{array}
\end{equation}
such that we have for the Poisson brackets of the functionals
$\, {\bf S} ({\bf X})$, $\, {\bf Q} ({\bf X})$,
$\, {\bar {\bf N}} ({\bf X})$:

\begin{equation}
\label{CanFormAvBrAnn}
\begin{array}{c}
\left\{ S^{\alpha} ({\bf X}) \, , \, S^{\beta} ({\bf Y}) \right\}
\,\, = \,\, 0     \\  \\
\left\{ S^{\alpha} ({\bf X}) \, , \, Q_{\beta} ({\bf Y}) \right\}
\,\, = \,\, \delta^{\alpha}_{\beta} \,\,
\delta ({\bf X} - {\bf Y}) \,\,\, , \,\,\,\,\,
\left\{ S^{\alpha} ({\bf X}) \, , \,
{\bar N}^{l}  ({\bf Y}) \right\}  \,\, = \,\, 0      \\  \\
\left\{ Q_{\alpha} ({\bf X}) \, , \, Q_{\beta} ({\bf Y}) \right\}
\,\, = \,\, 0 \,\,\, , \,\,\,\,\,
\left\{ Q_{\alpha} ({\bf X}) \, , \, 
{\bar N}^{l}  ({\bf Y}) \right\} \,\, = \,\, 0    \\  \\    
\left\{  {\bar N}^{l}  ({\bf X}) \, , \, 
{\bar N}^{k}  ({\bf Y}) \right\}
\,\,\,\,\, = \,\,\,\,\, g^{lk, p}
\,\, \delta_{X^{p}} ({\bf X} - {\bf Y}) 
\end{array}
\end{equation}
($g^{lk, p} \, = \, {\rm const}$).

}

\vspace{0.3cm}

 Proof.

 Let us consider the bracket (\ref{AveragedBracket}) in the
coordinates
$$\left( S^{1} ({\bf X}), \dots , S^{m} ({\bf X}), \,
{\tilde Q}_{1} ({\bf X}), \dots ,
{\tilde Q}_{m} ({\bf X}), \, N^{1} ({\bf X}), \dots ,
N^{s} ({\bf X}) \right) \,\,\, , $$
where it has the form (\ref{IntermedSQNBr}).

 Let us assume here without loss of generality that we have
$${\rm det} \, \left| \left| g^{lk, 1} \right| \right|
\,\,\, \neq \,\,\, 0 $$
for the pairwise Poisson brackets of the densities of annihilators.

 Let us consider now the Jacobi identities
\begin{multline*}
\left\{ \left\{ {\tilde Q}_{\alpha} ({\bf X}) \, , \,
N^{l} ({\bf Y}) \right\} \, , \, N^{k} ({\bf Z}) \right\}
\,\, - \,\, \left\{ \left\{ {\tilde Q}_{\alpha} ({\bf X}) \, , \,
N^{k} ({\bf Z}) \right\} \, , \, N^{l} ({\bf Y}) \right\}
\,\, -   \\
- \,\, \left\{ {\tilde Q}_{\alpha} ({\bf X}) \, , \,
\left\{ N^{l} ({\bf Y}) \, , \, N^{k} ({\bf Z}) \right\} \right\}
\,\,\, \equiv \,\,\, 0 
\end{multline*}

 It's not difficult to check that the identities above are 
equivalent to the following set of relations
\begin{equation}
\label{AgFirstRel}
{\partial A^{l, p}_{\alpha} \over \partial N^{r}} \,\,
g^{rk, q} \,\,\, - \,\,\,
{\partial A^{k, q}_{\alpha} \over \partial N^{r}} \,\,
g^{rl, p} \,\,\,\,\, = \,\,\,\,\, 0
\end{equation}

 Let us put now in (\ref{AgFirstRel}): 
$\,\, q \, = \, p \, = \, 1 \, . \, $  We get then
$${\partial A^{l, 1}_{\alpha} \over \partial N^{r}} \,\,
g^{rk, 1} \,\,\, = \,\,\,
{\partial A^{k, 1}_{\alpha} \over \partial N^{r}} \,\,
g^{rl, 1} $$

 Using the inverse tensor $\, g^{1}_{lk} \, $ we can write the
relation above in the equivalent form
$${\partial \over \partial N^{l}} \, \left( A^{r, 1}_{\alpha} \,
g^{1}_{rk} \right) \,\,\,\,\, = \,\,\,\,\,
{\partial \over \partial N^{k}} \, \left( A^{r, 1}_{\alpha} \,
g^{1}_{rl} \right) $$
which implies
$$A^{r, 1}_{\alpha} \, g^{1}_{rk} \,\,\,\,\, \equiv \,\,\,\,\,
{\partial f_{\alpha} \over \partial N^{k}} $$
or
$$A^{l, 1}_{\alpha} \,\,\,\,\, \equiv \,\,\,\,\,
{\partial f_{\alpha} \over \partial N^{k}} \,\,
g^{kl, 1} $$
for some functions 
$\, f_{\alpha} ({\bf S}_{X^{1}}, \dots , {\bf S}_{X^{d}}, \,
{\bf N} ) \, . $

 Let us just put now $\, q \, = \, 1 \, $  in relations
(\ref{AgFirstRel}). We immediately get then
$${\partial A^{l, p}_{\alpha} \over \partial N^{j}} 
\quad  =  \quad 
{\partial A^{k, 1}_{\alpha} \over \partial N^{r}} \,\,\,
g^{rl, p} \,\, g^{1}_{kj}  \quad  =  \quad
{\partial \over \partial N^{r}} \, \Big( A^{k, 1}_{\alpha} \,
g^{rl, p} \, g^{1}_{kj} \Big)  \quad  =  \quad
{\partial \over \partial N^{r}} \,
{\partial \over \partial N^{j}} \, \Big( f_{\alpha} \,
g^{rl, p} \Big)  \quad ,  \quad \quad   \forall p $$
 
 Thus, we can put
$$A^{l, p}_{\alpha}  \quad  \equiv  \quad
{\partial f_{\alpha} \over \partial N^{k}} \,\,
g^{kl, p}  \quad  +  \quad  \gamma^{l, p}_{\alpha} \left(
{\bf S}_{X^{1}}, \dots , {\bf S}_{X^{d}} \right)
\quad , \quad \quad  (p \geq 2) $$ 
for some functions 
$\, \gamma^{l, p}_{\alpha} \,
( {\bf S}_{X^{1}}, \dots , {\bf S}_{X^{d}} ) $.

 Let us put now
$$Q^{\prime}_{\alpha} ({\bf X})  \quad  =  \quad 
{\tilde Q}_{\alpha} ({\bf X})  \quad  -  \quad
f_{\alpha} \left( {\bf S}_{X^{1}}, \dots , {\bf S}_{X^{d}}, \,
{\bf N} \right) $$

 Easy to see that we have then:
$$ \left\{ Q^{\prime}_{\alpha} ({\bf X}) \, , \,
N^{l} ({\bf Y}) \right\}  \quad  =  \quad  \sum_{p=2}^{s} \,\,
\gamma^{l, p}_{\alpha} \, \left( {\bf S}_{X^{1}}, \dots , 
{\bf S}_{X^{d}} \right) \,\, \delta_{X^{p}} \,
({\bf X} - {\bf Y})  \quad  , $$
$$ \left\{ S^{\alpha} ({\bf X}) \, , \, 
Q^{\prime}_{\beta} ({\bf Y}) \right\}  \quad  =  \quad
\delta^{\alpha}_{\beta} \,\, \delta \, ({\bf X} - {\bf Y}) $$
for the new coordinates $\, Q^{\prime}_{\alpha} ({\bf X}) $.

 Consider now the Jacobi identities
\begin{multline}
\label{QprimeQprimeNJiden}
\left\{ Q^{\prime}_{\alpha} ({\bf X}) \, , \,
\left\{ Q^{\prime}_{\beta} ({\bf Y}) \, , \, 
N^{l} ({\bf Z}) \right\} \right\}
\,\,\, - \,\,\,
\left\{ Q^{\prime}_{\beta} ({\bf Y}) \, , \,
\left\{ Q^{\prime}_{\alpha} ({\bf X}) \, , \, 
N^{l} ({\bf Z}) \right\} \right\}  \quad  \equiv   \\
\equiv  \quad  \left\{ \left\{ 
Q^{\prime}_{\alpha} ({\bf X}) \, , \,
Q^{\prime}_{\beta} ({\bf Y}) \right\} \, , \,
N^{l} ({\bf Z}) \right\}
\end{multline}

 It's not difficult to check that from the identities
(\ref{QprimeQprimeNJiden}) and the conditions
$$\gamma^{l, 1}_{\alpha}  \,\,\,  \equiv  \,\,\,  0 
\quad  ,  \quad  \quad  
{\rm det} \, \left| \left| g^{kl, 1} \right| \right|
\,\,\, \neq \,\,\, 0 $$
we immediately get the relations
$${\delta \over \delta N^{k} ({\bf W})} \,\,
\left\{ Q^{\prime}_{\alpha} ({\bf X}) \, , \,
Q^{\prime}_{\beta} ({\bf Y}) \right\} \quad  \equiv  \quad  0 $$
which means
$$\left\{ Q^{\prime}_{\alpha} ({\bf X}) \, , \,
Q^{\prime}_{\beta} ({\bf Y}) \right\} \quad  =  \quad
\left\{ Q^{\prime}_{\alpha} ({\bf X}) \, , \,
Q^{\prime}_{\beta} ({\bf Y}) \right\} \left[ {\bf S} \right] $$

 Applying Theorem 2.1 to the set of the variables
$\, (S^{1} ({\bf X}), \dots , S^{m} ({\bf X}), \,
Q^{\prime}_{1} ({\bf X}), 
\dots , Q^{\prime}_{m} ({\bf X})) $,   \linebreak
we can define the variables
$$Q_{\alpha} ({\bf X})  \quad  =  \quad
Q^{\prime}_{\alpha} ({\bf X})  \quad  +  \quad
q^{\prime}_{\alpha} \left( {\bf S}_{X^{1}}, \dots , {\bf S}_{X^{d}}
\right) $$
satisfying the relations
$$\left\{ S^{\alpha} ({\bf X}) \, , \, Q_{\beta} ({\bf Y}) \right\}
\,\,\, = \,\,\, \delta^{\alpha}_{\beta} \,\,
\delta ({\bf X} - {\bf Y})  \quad  ,  \quad  \quad
\left\{ Q_{\alpha} ({\bf X}) \, , \, Q_{\beta} ({\bf Y}) \right\}
\,\,\, = \,\,\, 0  \quad  , $$
$$ \left\{ Q_{\alpha} ({\bf X}) \, , \,
N^{l} ({\bf Y}) \right\}  \quad  =  \quad  \sum_{p=2}^{s} \,\,
\gamma^{l, p}_{\alpha} \, \left( {\bf S}_{X^{1}}, \dots ,
{\bf S}_{X^{d}} \right) \,\, \delta_{X^{p}} \,
({\bf X} - {\bf Y}) $$

 Another corollary of the identities (\ref{QprimeQprimeNJiden})
is given by the relations
\begin{equation}
\label{gammaSderSkewSymm}
{\partial \gamma^{l, p}_{\alpha} \over \partial
S^{\beta}_{X^{q}}} \,\,\,  \equiv  \,\,\, - \,
{\partial \gamma^{l, p}_{\beta} \over \partial
S^{\alpha}_{X^{q}}}  \quad  ,  \quad  \quad
{\partial \gamma^{l, p}_{\alpha} \over \partial
S^{\beta}_{X^{q}}} \,\,\,  \equiv  \,\,\, - \,
{\partial \gamma^{l, q}_{\alpha} \over \partial
S^{\beta}_{X^{p}}}  
\end{equation}
($ p \, = \, 2, \dots, s \, , \,\,  
q \, = \, 1, \dots, s \, $), $\, $  and
$$\left[ {\partial \gamma^{l, p}_{\alpha} \over \partial
S^{\beta}_{X^{q}}} \right]_{X^{q}} \,\,\,  \equiv  \,\,\, 0 $$

 Relations (\ref{gammaSderSkewSymm}) give, in particular,
the relations
$${\partial \gamma^{l, p}_{\alpha} \over \partial
S^{\beta}_{X^{q}}}  \,\,\,  \equiv  \,\,\,
{\partial \gamma^{l, q}_{\beta} \over \partial
S^{\alpha}_{X^{p}}} $$
which implies the relations
$$ \gamma^{l, p}_{\alpha}  \quad  \equiv  \quad
{\partial \over \partial S^{\alpha}_{X^{p}}} \,\,\,
g^{l} \left( {\bf S}_{X^{2}}, \dots , {\bf S}_{X^{d}} \right)
\quad  ,  \quad  \quad  \quad   (p \geq 2) $$
for some functions
$\, g^{l} \, ({\bf S}_{X^{2}}, \dots , {\bf S}_{X^{d}}) $.

 Let us put now
$${\bar N}^{l} ({\bf X})  \quad  =  \quad   
N^{l} ({\bf X})  \quad  -  \quad
g^{l} \left( {\bf S}_{X^{2}}, \dots , {\bf S}_{X^{d}} \right) $$

 Using again the relations (\ref{gammaSderSkewSymm}) it's not
difficult to show then that we obtain
$$ \left\{ Q_{\alpha} ({\bf X}) \, , \,
{\bar N}^{l} ({\bf Y}) \right\}  \,\,\,  =  \,\,\,  0 $$ 
for the variables $\, {\bar N}^{l} ({\bf Y}) $.

 Finally, we get the variables
$\, ( {\bf S} ({\bf X}), \, {\bf Q} ({\bf X}), \,
{\bar {\bf N}} ({\bf X}) ) \, $ satisfying all the relations
(\ref{CanFormAvBrAnn}).

{\hfill Theorem 3.1 is proved.}

\vspace{0.5cm}

 At the end, let us say that in general the separation of the 
variables $\, ({\bf S} ({\bf X}) , \, {\bf Q} ({\bf X}) ) \, $ 
and $\, {\bf N} ({\bf X}) \, $ into two independent brackets by 
a transformation of the form (\ref{UQNTrans}) is impossible for 
the bracket (\ref{AveragedBracket}). As an example, let us
consider the bracket (\ref{AveragedBracket}) which has in the
variables 
$$\left( S ({\bf X}), \, {\tilde Q}  ({\bf X}), \,
{\tilde N}^{1}  ({\bf X}), \, {\tilde N}^{2}  ({\bf X})
\right) $$
the following form
\begin{equation}
\label{ExampleBracket}
\begin{array}{c}
\left\{ S ({\bf X}) \, , \, S ({\bf Y}) \right\}
\,\,\, = \,\,\, 0     \quad  ,  \quad \quad
\left\{ S ({\bf X}) \, , \,  {\tilde Q} ({\bf Y}) \right\}
\,\,\, = \,\,\, \delta \, ({\bf X} - {\bf Y})   
\quad  ,  \quad \quad  
\left\{ {\tilde Q} ({\bf X}) \, , \, {\tilde Q} ({\bf Y})
\right\} \,\,\, = \,\,\, 0  \\  \\
\left\{ S ({\bf X}) \, , \, {\tilde N}^{1} ({\bf Y}) \right\}
\quad  =  \quad 
\left\{ S({\bf X}) \, , \, {\tilde N}^{2} ({\bf Y}) \right\}
\quad  =  \quad   0     \\  \\
\left\{ {\tilde Q} ({\bf X}) \, , \, {\tilde N}^{1} ({\bf Y}) 
\right\}  \quad  =  \quad  {\tilde N}^{2} ({\bf X}) \,\,
\delta_{X^{2}} ({\bf X} - {\bf Y})  \quad  ,  \quad \quad
\left\{ {\tilde Q} ({\bf X}) \, , \, {\tilde N}^{2} ({\bf Y})
\right\}  \,\,\, = \,\,\,  0     \\  \\
\left\{  \begin{pmatrix}  {\tilde N}^{1} ({\bf X})   \\  
{\tilde N}^{2} ({\bf X}) \end{pmatrix} \,\, , \,\,
\left( {\tilde N}^{1} ({\bf Y}) \,\,\,
{\tilde N}^{2} ({\bf Y}) \right)  \right\} \quad  =  \quad 
\begin{pmatrix} S_{X^{2}}  &  1   \\  
1  &  0  \end{pmatrix} \,\, 
\delta_{X^{1}} ({\bf X} - {\bf Y})  \quad  -  \quad 
\begin{pmatrix} S_{X^{1}}  &  0   \\
0  &  0  \end{pmatrix} \,\,
\delta_{X^{2}} ({\bf X} - {\bf Y})  
\end{array}
\end{equation}
($ \, {\bf X} \, = \, (X^{1}, \, X^{2}) \, $).

 It can be checked by direct calculation that the bracket
(\ref{ExampleBracket}) is skew-symmetric and satisfies the
Jacobi identity. However, it's not difficult to check that
no transformation
$$\left( {\tilde Q}  ({\bf X}), \,
{\tilde N}^{1}  ({\bf X}), \, {\tilde N}^{2}  ({\bf X})
\right)  \quad  \rightarrow  \quad  
\Big( Q  ({\bf X}), \, N^{1}  ({\bf X}), \, N^{2}  ({\bf X})
\Big) $$
where
$$Q ({\bf X}) \,\,\, = \,\,\, {\tilde Q} ({\bf X}) \,\, + \,\,
{\tilde q} \, \left( S_{X^{1}}, \, S_{X^{2}}, \, 
{\tilde N}^{1} ({\bf X}), \, {\tilde N}^{2} ({\bf X}) \right) 
\,\,\,  ,  $$
$$ N^{l} ({\bf X}) \,\,\, = \,\,\, N^{l}
\left( S_{X^{1}}, \, S_{X^{2}}, \,
{\tilde N}^{1} ({\bf X}), \, {\tilde N}^{2} ({\bf X}) \right) $$
can give the relations
\begin{equation}
\label{NewVarZeroRel}
\left\{ Q ({\bf X}) \, , \, N^{1} ({\bf Y})
\right\}  \quad  =  \quad 
\left\{ Q ({\bf X}) \, , \, N^{2} ({\bf Y})
\right\}  \quad  =  \quad  0 
\end{equation}

 Indeed, according to the Jacobi identities
\begin{multline*}
\left\{ \left\{ N^{l} ({\bf X}) \, , \,
N^{k} ({\bf Y}) \right\} \, , \, Q ({\bf Z}) \right\}
\,\, + \,\, \left\{ \left\{ Q ({\bf Z}) \, , \,
N^{l} ({\bf X}) \right\} \, , \, N^{k} ({\bf Y}) \right\}
\,\, +   \\
+ \,\, \left\{ \left\{ N^{k} ({\bf Y}) \, , \,
Q ({\bf Z}) \right\} \, , \, N^{l} ({\bf X}) \right\}
\,\,\, \equiv \,\,\, 0   \,\,\, ,
\end{multline*}
the fulfillment of the conditions (\ref{NewVarZeroRel}) would
imply the relations
$${\delta \over \delta S ({\bf Z})} \,
\left\{ N^{l} ({\bf X}) \, , \, N^{k} ({\bf Y}) \right\} 
\,\,\, \equiv \,\,\, 0   \,\,\, ,  $$
which means the independence of the brackets
$\, \{ N^{l} ({\bf X}) \, , \, N^{k} ({\bf Y}) \} \, $
on the variables $\, [ S ({\bf Z}) ] $.

 On the other hand, the Poisson brackets
$\, \{ N^{l} ({\bf X}) \, , \, N^{k} ({\bf Y}) \} \, $
can be represented in the form
\begin{multline*}
\left\{  \begin{pmatrix}  N^{1} ({\bf X})   \\
N^{2} ({\bf X}) \end{pmatrix} \,\, , \,\,
\left( N^{1} ({\bf Y}) \,\,\,
N^{2} ({\bf Y}) \right)  \right\} \quad  =   \\
= \,\,\,  \begin{pmatrix} 
\partial N^{1} / \partial {\tilde N}^{1}  &
\partial N^{1} / \partial {\tilde N}^{2}  \\
\partial N^{2} / \partial {\tilde N}^{1}  & 
\partial N^{2} / \partial {\tilde N}^{2}  
\end{pmatrix} \,\,\, \times \,\,\, \left[
\begin{pmatrix} S_{X^{2}}  &  1   \\
1  &  0  \end{pmatrix} 
\begin{pmatrix}
\partial N^{1} / \partial {\tilde N}^{1}  &
\partial N^{2} / \partial {\tilde N}^{1}  \\
\partial N^{1} / \partial {\tilde N}^{2}  &
\partial N^{2} / \partial {\tilde N}^{2} 
\end{pmatrix} \,\,\, 
\delta_{X^{1}} ({\bf X} - {\bf Y})  \quad  -   \right.  \\
-  \quad  \begin{pmatrix} S_{X^{1}}  &  0   \\
0  &  0  \end{pmatrix}
\begin{pmatrix}
\partial N^{1} / \partial {\tilde N}^{1}  &
\partial N^{2} / \partial {\tilde N}^{1}  \\
\partial N^{1} / \partial {\tilde N}^{2}  &
\partial N^{2} / \partial {\tilde N}^{2}
\end{pmatrix} \,\,\,
\delta_{X^{2}} ({\bf X} - {\bf Y})  \quad   +   \\
+  \quad  \begin{pmatrix} S_{X^{2}}  &  1   \\
1  &  0  \end{pmatrix}
\begin{pmatrix}
\partial N^{1} / \partial {\tilde N}^{1}  &
\partial N^{2} / \partial {\tilde N}^{1}  \\
\partial N^{1} / \partial {\tilde N}^{2}  &
\partial N^{2} / \partial {\tilde N}^{2}
\end{pmatrix}_{X^{1}} \,\,\,
\delta \, ({\bf X} - {\bf Y}) \quad  -  \\
\left.  -  \quad  \begin{pmatrix} S_{X^{1}}  &  0   \\
0  &  0  \end{pmatrix}
\begin{pmatrix}
\partial N^{1} / \partial {\tilde N}^{1}  &
\partial N^{2} / \partial {\tilde N}^{1}  \\
\partial N^{1} / \partial {\tilde N}^{2}  &
\partial N^{2} / \partial {\tilde N}^{2}
\end{pmatrix}_{X^{2}} \,\,\,
\delta \, ({\bf X} - {\bf Y})  \right]
\end{multline*}

 We can see then, that the absence of the terms containing the 
derivatives $\, S_{X^{1} X^{1}} \, $ and $\, S_{X^{1} X^{2}} \, $ 
in the brackets 
$\, \{ N^{l} ({\bf X}) \, , \, N^{k} ({\bf Y}) \} \, $
requires, in particular, the relations
$${\partial^{2} N^{l} \over \partial {\tilde N}^{1} \,
\partial S_{X^{1}} } \,\, = \,\, 0
\quad  ,  \quad \quad 
{\partial^{2} N^{l} \over \partial {\tilde N}^{1} \,
\partial S_{X^{2}} } \,\, = \,\, 0
\quad  ,  \quad \quad
l \, = \, 1, \, 2  \quad  , $$
which means in fact
\begin{equation}
\label{FormofNNTrans}
N^{l} \, \Big( S_{X^{1}}, \, S_{X^{2}}, \, {\tilde N}^{1}, \,
{\tilde N}^{2} \Big)  \quad  \equiv  \quad  
N^{\prime l} \, \Big( {\tilde N}^{1}, \, {\tilde N}^{2} \Big) 
\,\,\, + \,\, N^{\prime\prime l} \, \Big( S_{X^{1}}, \, 
S_{X^{2}}, \, {\tilde N}^{2} \Big) 
\end{equation}

 On the other hand, it's not difficult to see that the 
transformations (\ref{FormofNNTrans}) can not transform the
metric
$$g^{lk, 2} \left( S_{\bf X} \right)  \quad  =  \quad  
- \,\, \begin{pmatrix} S_{X^{1}}  &  0   \\
0  &  0  \end{pmatrix}  $$
into a form, independent on $\, S_{X^{1}} $. So, we get now
our statement.

\vspace{0.3cm}

 The work was partially supported by Grant RFBR 
No. 13-01-12469-ofi-m-2013.

\end{document}